\documentclass[12pt]{iopart}
\usepackage{graphicx}

\begin{document}

\bibliographystyle{unsrt}

\title{High-precision measurement of hyperfine structure in the
$D$ lines of alkali atoms}
\author{Dipankar Das and Vasant Natarajan}
\address{Department of Physics, Indian Institute of Science,
Bangalore 560 012, INDIA}
\ead{vasant@physics.iisc.ernet.in}

\begin{abstract}
We have measured hyperfine structure in the first-excited
$P$ state ($D$ lines) of all the naturally-occurring alkali
atoms. We use high-resolution laser spectroscopy to resolve
hyperfine transitions, and measure intervals by locking the
frequency shift produced by an acousto-optic modulator to
the difference between two transitions. In most cases, the
hyperfine coupling constants derived from our measurements
improve previous values significantly.
\end{abstract}



\maketitle

\section{Introduction}

High-precision measurement of hyperfine structure in the
excited states of alkali atoms provides a stringent testing
ground for state-of-the-art atomic calculations based on
the best wavefunctions \cite{SJD99}, because it is
sensitive to effects such as core polarization and electron
correlation. The $D_2$ line of alkali atoms is routinely
used in laser cooling and Bose-Einstein condensation
experiments. These ultracold atoms are important in
experiments ranging from fundamental physics issues such as
high-resolution spectroscopy, measurement of fundamental
constants, and the study of cold collisions, to
applications in atomic clocks and inertial sensors. Heavy
alkali atoms are also important in studies of atomic parity
violation and the search for a permanent atomic electric
dipole moment. In these experiments, accurate calculations
are needed to properly interpret the experimental results.
In addition, hyperfine structure provides important
information about the structure of the nucleus, i.e.\ the
multipole moments of its charge and current distribution
\cite{AIV77}.

Hyperfine structure has been measured with high accuracy in
the ground state of alkali atoms using atomic beam magnetic
resonance techniques. Thus the hyperfine constants are
known with a relative precision of better than $10^{-9}$.
Indeed, this high precision of the measurements has led to
the definition of the SI unit of time in terms of the
ground hyperfine interval in Cs. However, measurement of
hyperfine structure in the excited state is often limited
by the linewidth of the transition, which is about 5--10
MHz for the $D$ lines of alkali atoms. The advent of narrow
linewidth tunable lasers (ring dye lasers and
frequency-stabilized diode lasers) and high-resolution
Doppler-free spectroscopy techniques (such as
saturated-absorption spectroscopy) has allowed individual
hyperfine transitions to be resolved. But there is still an
issue of frequency calibration and linearity of the laser
scan axis. We have solved this problem by using an
acousto-optic modulator (AOM), driven by a easily-measured
radio frequency source, to span the difference between
hyperfine transitions. We are thus able to measure
hyperfine intervals in the $D$ lines of alkali atoms with a
typical precision of 1--2 parts in 1000 of the linewidth.
In earlier work, we had reported measurement of hyperfine
structure in a few alkali atoms. In the current work, we
present comprehensive results on all the
naturally-occurring alkali atoms. The results, in most
cases, are consistent with previously published values but
represent a significant improvement in precision.

Let us first consider the definition of the hyperfine
constants. The basis states used for defining hyperfine
structure are the $| J,I,F,m_F \rangle$ states, where $J$,
$I$, and $F$ are the quantum numbers associated with,
respectively, the electronic angular momentum ${\bf J}$,
the nuclear spin ${\bf I}$, and the total angular momentum
${\bf F=I+J}$. In the approximation that $J$ and $I$ are
good quantum numbers, the hyperfine interaction energy
$W_F$ is given by,
\[
W_F = \frac{1}{2} hAK + hB
\frac{\frac{3}{2}K(K+1)-2I(I+1)J(J+1)}{2I(2I-1)2J(2J-1)} \;
,
\]
where $K=F(F+1)-I(I+1)-J(J+1)$, $A$ is the magnetic-dipole
coupling constant, and $B$ is the electric-quadrupole
coupling constant.

The first excited $P$ state of alkali atoms splits into two
states, $P_{1/2}$ and $P_{3/2}$, due to spin-orbit
interaction (fine-structure splitting). Of these, the
$P_{1/2}$ state ($D_1$ line) further splits into two
hyperfine levels with energy shifts determined by the
magnetic-dipole interaction. The electric-quadrupole
interaction is present only for $I,J \geq 1$, and hence
does not affect this state. On the other hand, the
$P_{3/2}$ state ($D_2$ line) splits into four hyperfine
levels with energy shifts determined by both interactions.
In the case of $^{133}$Cs, we will see later that there is
evidence of a small additional contribution from a
magnetic-octupole hyperfine interaction.

\section{Experimental technique}
The experimental schematic to measure hyperfine intervals
is shown in Fig.\ 1, and is essentially the same for all
atoms. The output from a frequency-stabilized tunable laser
is split into two parts. The first part goes into an atomic
spectrometer, the signal from which is used to lock the
laser to a particular hyperfine transition. The second part
is frequency shifted through an AOM and then sent into a
second atomic spectrometer. The frequency of the AOM is
adjusted so that the shifted beam is on a neighboring
hyperfine transition. The error signal from this peak is
fed back to the AOM driver to lock its frequency. Thus the
AOM frequency gives a direct measurement of the hyperfine
interval. For intervals that are too small or too large to
be measured with a single AOM, we use an additional AOM
with a fixed frequency offset.

The $D$ lines in sodium are in the visible (at 589 nm),
hence the tunable laser is a ring dye laser (Coherent
699-21). The laser is frequency stabilized to a reference
cavity that gives it an instantaneous linewidth of 1 MHz.
The $D$ lines in all the other atoms are in the near
infrared which can be accessed using diode lasers. The
diode laser is frequency stabilized using optical feedback
from a piezo-mounted grating so that the resulting
linewidth is about 500 kHz. The linewidth is reduced
considerably when the laser is locked. For locking, the
laser is frequency modulated at $f=20$ kHz, and the signal
from the spectrometers is demodulated at $3f$ to generate
the error signals. Such third-harmonic detection provides
narrow dispersive signals that are insensitive to intensity
fluctuations, to peak pulling from neighboring transitions,
or to any residual Doppler profile in the spectrum
\cite{WAL72}.

In the case of Li, the spectroscopy is done using a
collimated atomic beam. The atoms are excited by a
perpendicular laser beam, and the resulting Doppler-free
fluorescence signal is measured with a photomultiplier tube
(PMT). In all the other atoms, we use absorption
spectroscopy of a probe beam through a vapor cell. The cell
is either at room temperature (Rb and Cs) or heated to
$75^\circ$C (Na and K), to have an atomic density of about
$10^9$ cm$^{-3}$. In order to get Doppler-free absorption
spectra, we use one of two techniques. The first technique
is the normal saturated-absorption spectroscopy (SAS)
\cite{DEM82}, where a strong counter-propagating pump beam
decreases the absorption of the weak probe beam on
resonance. As is well known, the SAS technique produces
additional crossover resonances, which occur exactly midway
between two hyperfine peaks. For closely-spaced levels, the
large crossover resonances often swamp the true peaks. This
is a problem especially in the $D_2$ lines of Na and K,
where the level spacing is of the order of one or two
natural linewidths.

To overcome this problem, we have recently developed an
alternate technique of high-resolution spectroscopy, which
we call coherent-control spectroscopy (CCS)
\cite{BAN03,DAN05,DPW06}. This technique uses
co-propagating control and probe beams, and does not
produce crossover resonances. The probe beam is locked on
to a hyperfine transition, while the control beam is
scanned across a neighboring transition. When the control
comes into resonance with a transition coupling to the same
ground level, it ``coherently'' reduces probe absorption
through a process similar to electromagnetically induced
transparency. Since the probe is locked on a transition, it
addresses only the zero-velocity atoms, and the resultant
lineshape remains Doppler-free even in hot vapor.

We have also done a density-matrix analysis of the
effective three-level atom in the presence of the probe and
control beams. The calculated lineshape, after accounting
for thermal averaging in hot vapor, describes the measured
spectrum quite well \cite{DAN05}. Apart from the ability to
resolve closely-spaced levels, another important advantage
of the CCS technique for measuring hyperfine intervals is
that the measured value is completely insensitive to
detuning of the probe beam. Any detuning of the probe from
resonance would imply that it is resonant with a non-zero
velocity group. Since the control beam co-propagates with
the probe beam, the spectrum will show peaks only when the
control beam comes into resonance with the same non-zero
velocity group, which happens only when the AOM shift
matches the hyperfine interval. Thus, as the probe beam is
detuned from resonance, the manifold of peaks
(corresponding to the different hyperfine levels) will
shift within the Doppler profile, but their relative
separation will remain the same.

\section{Error Analysis}
The different sources of error in the measurement have been
discussed extensively in our earlier publications, and are
reviewed here for completeness.

\subsection{Statistical errors}
The primary sources of statistical error are the
fluctuations in the lock point of the laser and the AOM. To
minimize these errors, we use an integration time of 10 s
in the frequency counter during each measurement of the AOM
frequency. Then we take an average of 35--40 measurements
for a given transition, and repeat the set several times.
The resulting standard deviation (scatter) in the data set
is about 10 kHz, from which we take the statistical error
in the mean to be $\sim 2$ kHz. The timebase in the
frequency counter used for measuring the AOM frequency has
a stability of better than $10^{-6}$, which translates to a
negligible error of $<0.1$ kHz in the frequency
measurement.

\subsection{Systematic errors}
Systematic errors can occur if there are systematic shifts
in the lock points of the laser and the AOM. This can arise
due to one of the following reasons.
\begin{enumerate}
\item[(i)] {\it Radiation-pressure effects.} Radiation
pressure causes velocity redistribution of the atoms in the
vapor cell. In the SAS technique, the opposite Doppler
shifts for the counter-propagating beams can result in
asymmetry of the observed lineshape \cite{GRM89}. We
minimize these effects by using beam intensities that are
much smaller than the saturation intensity. In the CCS
technique, radiation pressure effects are less important
because the Doppler shift will be the same for both beams
and will not affect the hyperfine interval, similar to how
the interval is insensitive to any detuning of the probe
from resonance.

\item[(ii)] {\it Effect of stray magnetic fields.} The primary
effect of a magnetic field is to split the Zeeman sublevels
and broaden the line without affecting the line center.
However, line shifts can occur if there is asymmetric
optical pumping into Zeeman sublevels. For a transition
$|F,m_F \rangle \rightarrow |F',m_{F'} \rangle$, the
systematic shift of the line center is $\mu_B(g_{F'}m_{F'}
- g_Fm_F)B$, where $\mu_B=1.4$ MHz/G is the Bohr magneton,
$g$'s denote the Land\'e $g$ factors of the two levels, and
$B$ is the magnetic field. The selection rule for dipole
transitions is $\Delta m = 0,\pm 1$, depending on the
direction of the magnetic field and the polarization of the
light. Thus, if the beams are linearly polarized, there
will be no asymmetric driving and the line center will not
be shifted. We therefore minimize this error in two ways.
First, we use polarizing cubes to ensure that the beams
have near-perfect linear polarization. Second, we use a
special magnetic shield \cite{Conetic} around the cells to
minimize the field. The residual field (measured with a
3-axis fluxgate magnetometer) is below 5 mG with one layer
of shielding, and further reduces to 1 mG with two layers.

\item[(iii)] {\it Phase shifts in the feedback loop.} We check
for this error by replacing the AOM with two identical
AOMs, and adjusting them so that they produce opposite
frequency offsets. With the laser locked to a given
hyperfine transition, the first AOM then produces a fixed
frequency offset which is compensated by the second AOM.
Thus the same hyperfine transition is used for locking in
both spectrometers. Under these conditions, the second AOM
should lock to the fixed frequency of the first AOM, with
any error arising solely due to phase-shift errors. We find
that the second AOM tracks the frequency of the first AOM
to within 1 kHz.

\item[(iv)] {\it Shifts due to collisions.} To first order,
collisional shifts are the same for different hyperfine
levels, and hence do not affect the interval. Small
differential shifts of the interval have been studied
carefully in the ground state of Cs, due to its importance
in atomic clocks. However, the size of the shift is in the
mHz range \cite{SMB02}. We have also studied shifts in the
excited state of Cs by repeating the measurements with a
heated vapour cell. An increase from 25$^\circ$C to
45$^\circ$C increases the Cs density by a factor of 5, and
we find that the measured values shift by less than 5 kHz.
We therefore estimate that the maximum shift due to
collisions is 5 kHz.

\end{enumerate}

The sizes of the various sources of error are listed in
Table \ref{t1}. The size of the Zeeman shift is calculated
in each atom by considering all allowed combinations of
$m_F$ and $m_{F'}$ and taking the one with the largest
shift, assuming atoms get completely optically pumped into
these sublevels.

As mentioned in point (i) above, the peak center can be
shifted due to radiation-pressure effects. Similarly, from
point (ii), the center can be shifted if there is {\it
asymmetric} pumping into the Zeeman sublevels in the
presence of a residual magnetic field. Asymmetric optical
pumping can occur if the beam polarizations are not
perfectly linear, for example due to imperfections in the
cubes or birefringence at the cell windows. Finally, there
can be a small AC Stark shift of the transition due to
nearby hyperfine levels from which the laser is detuned.
From an experimental point of view, all these effects will
change with laser power. Therefore, we can check for these
errors by repeating the measurements at different values of
power, and extrapolate to zero power if necessary.


\section{Lithium}

We now consider the hyperfine measurements in each atom in
turn. We present a detailed description of the measurement
in Li for several reasons. First, Li is the only atom in
which we use an atomic beam for spectroscopy. Therefore, it
is slightly different from the vapor-cell technique used in
other atoms, which has been described in detail in our
earlier publications. Second, we give a complete list of
the hyperfine intervals measured at different powers. As
mentioned above, this method is used to check for
intensity-dependent errors, and we just present a summary
of the results in the other atoms. Finally, the small
ground hyperfine interval in Li offers a unique possibility
of checking the reliability of our technique, which is not
easy in the other atoms.

Lithium has two stable isotopes: $^{6}$Li with $I=1$ and
natural abundance of 7.3\%, and $^{7}$Li with $I=3/2$ and
natural abundance of 92.7\%. Li spectroscopy was done using
a collimated atomic beam generated by heating a stainless
steel oven containing Li metal to 300$^\circ$C. The oven
was placed inside a vacuum chamber maintained at a pressure
below $5 \times 10^{-8}$ torr with a 20 l/s ion pump. The
atomic beam was excited with a perpendicular laser beam
generated from a home-built diode laser system operating at
670 nm. The laser beam was linearly polarized and the
interaction region was shielded with a two-layer magnetic
shield. The resulting Doppler-free fluorescence signal was
detected by a PMT (Hamamatsu R928).

We obtained spectra from the $D$ lines of both isotopes. In
the $D_1$ line, the individual hyperfine transitions are
well resolved and we can lock to each peak separately.
However, in the $D_2$ line, the different transitions are
only partially resolved because the level spacing is less
than the natural linewidth. In recent work where we
measured the absolute frequencies of these transitions
\cite{DAN07}, we showed that a multipeak fit to the
partially-resolved spectrum yields the hyperfine constants
in the $2P_{3/2}$ state with a precision of about 50 kHz.
This is less than the precision with which they are known
from earlier experiments. Therefore, in this work we report
hyperfine structure only in the $2P_{1/2}$ state.

\subsection{$D_1$ line}

\begin{figure}
\hspace*{2cm}
\resizebox{0.43\columnwidth}{!}{\includegraphics{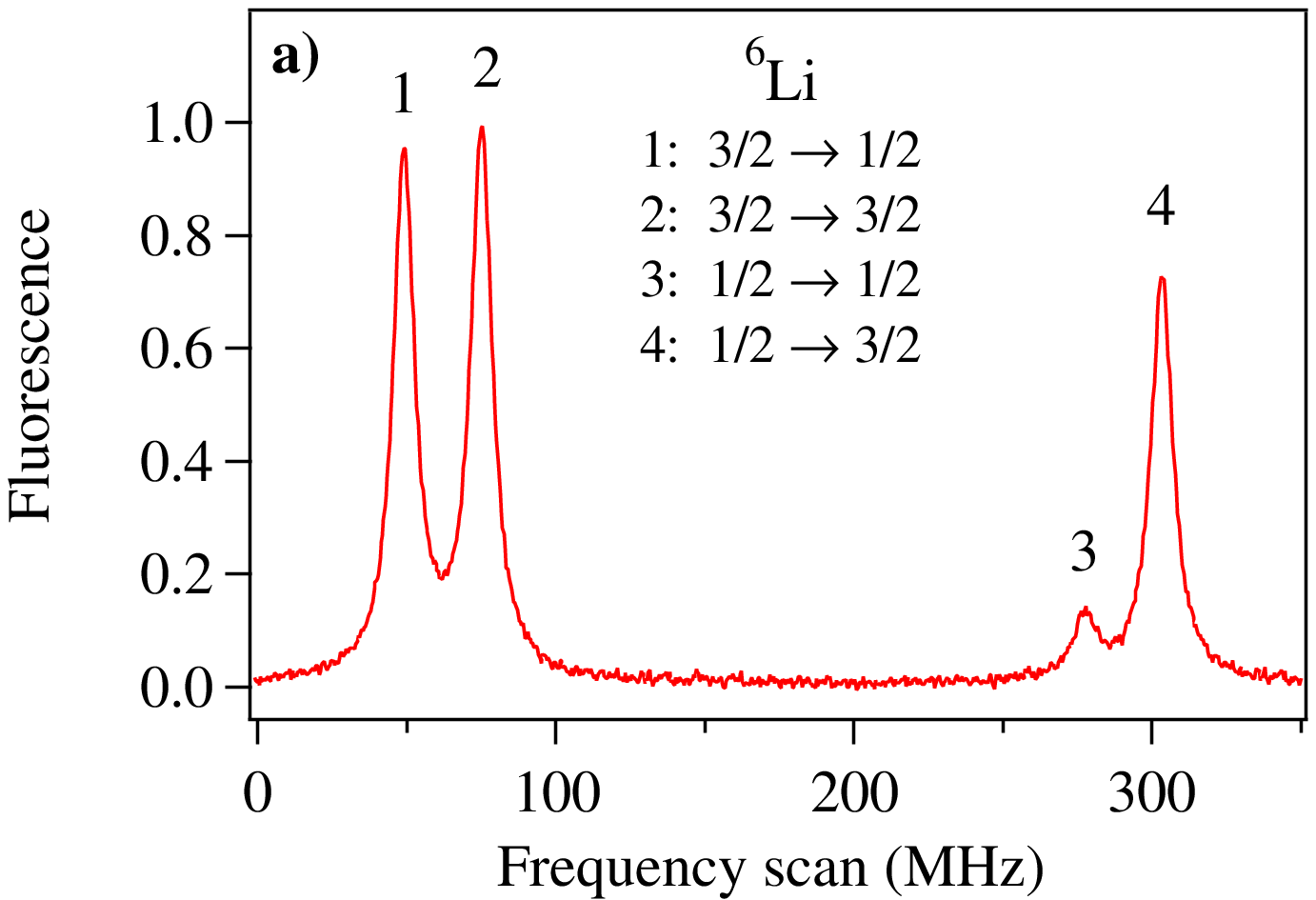}}
\resizebox{0.43\columnwidth}{!}{\includegraphics{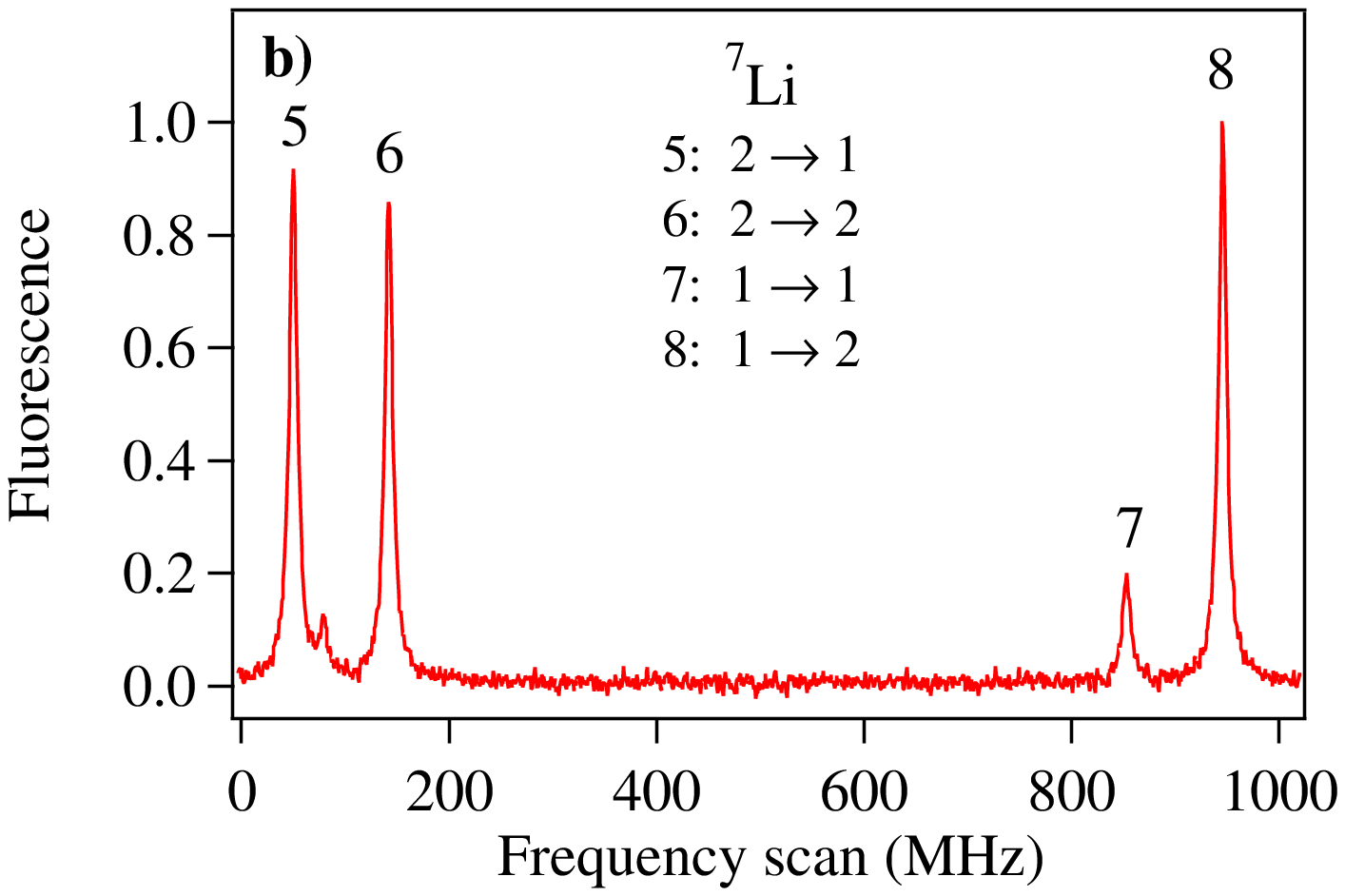}}\\
\hspace*{2cm}
\resizebox{0.43\columnwidth}{!}{\includegraphics{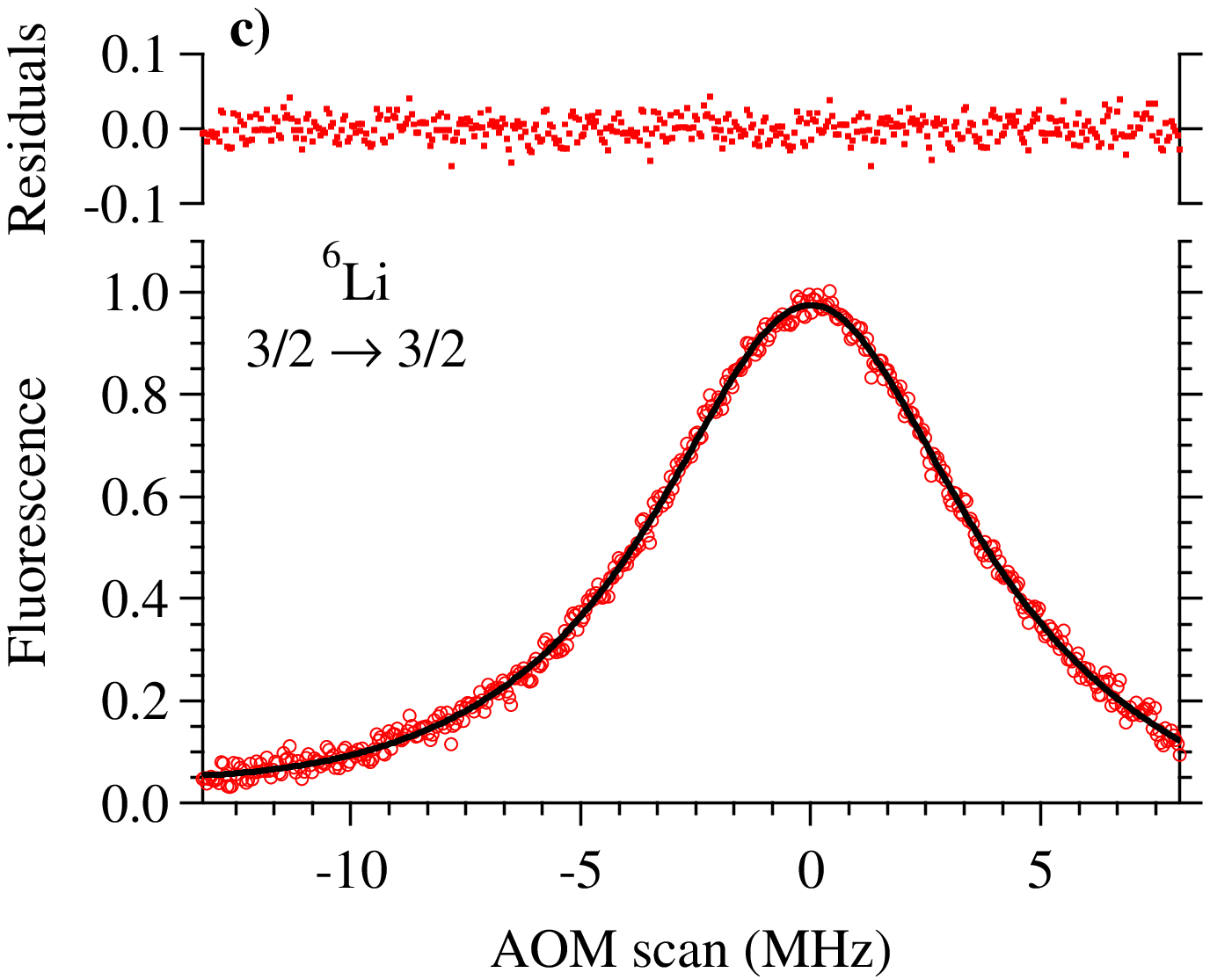}}
\resizebox{0.43\columnwidth}{!}{\includegraphics{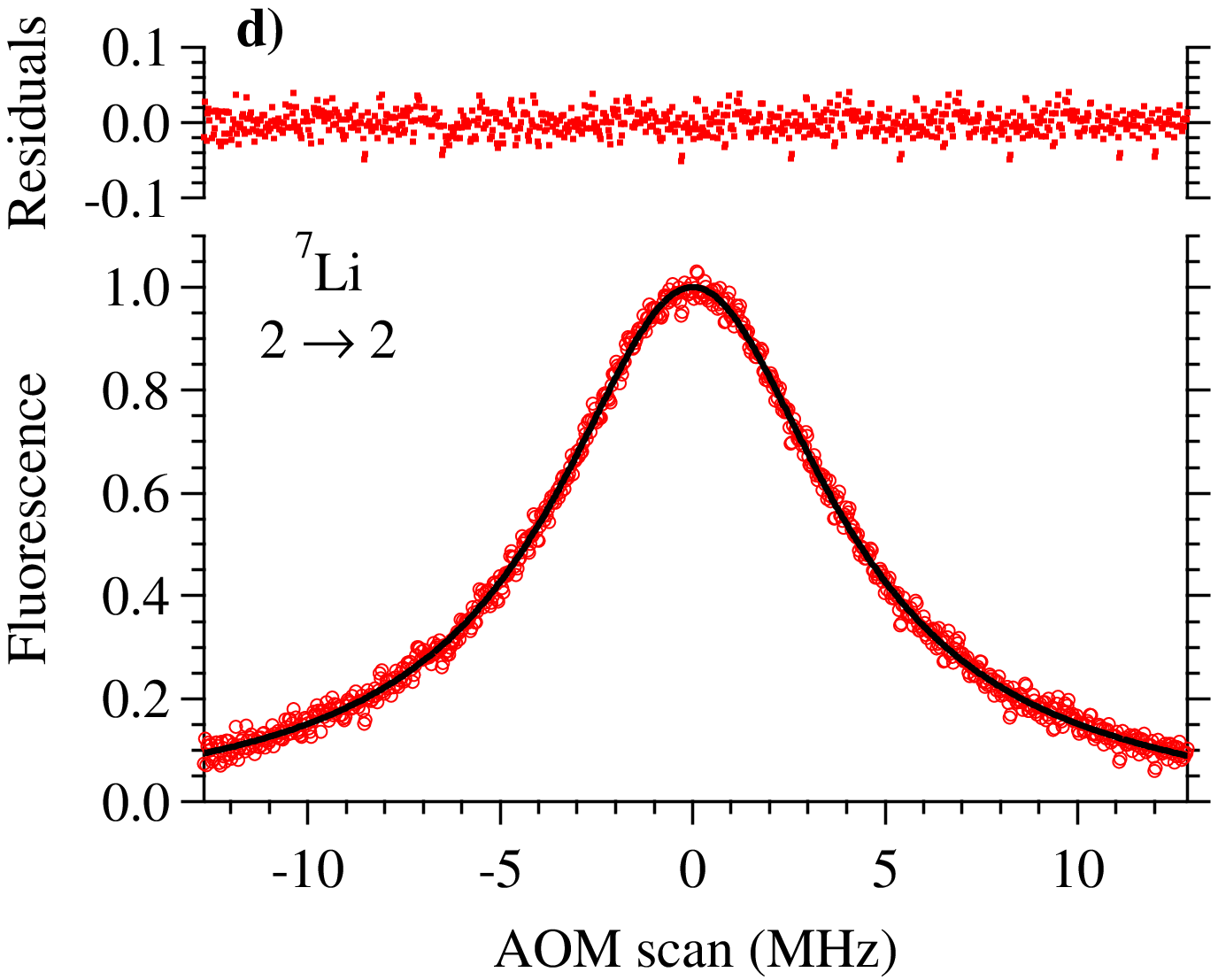}}
\caption{(Color online) $D_1$ line spectra in the two
isotopes of Li are shown in (a) and (b). The different
hyperfine transitions are clearly resolved, and labelled as
shown. There is a large increase in the gain of the PMT for
(a) compared to (b) to account for the low abundance of
$^6$Li. The small peak between peaks 5 and 6 is from the
nearby $D_2$ line in $^6$Li. Close-ups of peaks 2 and 6
(normalized) are shown in (c) and (d), respectively. The
circles are the measured spectra and the solid lines are
Lorentzian fits. The fit residuals give an idea of the
signal-to-noise ratio.} \label{f2a}
\end{figure}

Typical spectra for the $D_1$ line in the two isotopes of
Li are shown in Figs.\ \ref{f2a}(a) and (b). The four
hyperfine transitions in each isotope are clearly resolved.
Close-up scans (normalized) of a representative peak in
each isotope are shown in Figs.\ \ref{f2a}(c) and (d),
respectively. The solid lines are Lorentzian fits showing
the excellent fit with featureless residuals. The size of
the residuals gives an idea of the overall signal to noise
ratio. The fit linewidth is $9.0 \pm 0.25$ MHz (compared to
the natural linewidth of 5.87 MHz), indicating that the
atomic beam is well collimated because the spread in
transverse velocity increases the total linewidth by only
about 50\%. The laser linewidth does not contribute
significantly to this broadening.

An important consideration in atomic beam experiments is
the systematic Doppler shift of line center that can occur
if the laser beam is not perfectly orthogonal. In the
recent work on absolute frequency measurements of the $D$
lines \cite{DAN07}, we used the same spectrometer to
measure the frequencies with laser beams traversing the
atomic beam in opposite directions. Since the Doppler shift
is opposite for the two cases, their difference gives a
measure of the misalignment from perpendicularity. Our
measurements show that the Doppler shift of the transition
frequency is only 150 kHz, corresponding to a misalignment
angle of 0.1 mrad. For the hyperfine interval, this implies
a negligible error of less than 1 Hz.

In order to measure the hyperfine intervals, both the
unshifted laser beam and the AOM-shifted beam were sent
across the atomic beam. The diode laser and the AOM were
modulated at different frequencies, and the signal from the
PMT was independently demodulated at the two frequencies.
The first error signal was used to lock the laser on a
given hyperfine transition, while the second error signal
was use to lock the AOM to the hyperfine interval.

A stringent check on our error estimate in Table \ref{t1}
is to use our technique to measure the ground hyperfine
interval. As mentioned in the Introduction, this interval
is already known in all alkali atoms with $<10^{-9}$
relative precision. However, our technique requires the
interval to be accessible with an AOM, and this is possible
only in $^{6,7}$Li, where the intervals are about 200 MHz
and 800 MHz respectively. Among the different sources of
error in this measurement considered before, the effect of
collisions is negligible in an atomic beam. We can also
check for the intensity-dependent errors by varying the
laser power. The results of ground hyperfine measurements
at three powers are shown in Table \ref{t2}. The beam size
is such that the peak intensity (at beam center) for the
highest power of 42 $\mu$W is 0.84 mW/cm$^2$, which is
still smaller than the saturation intensity of 2.51
mW/cm$^2$.

\begin{figure}
\hspace*{2cm}
\resizebox{0.43\columnwidth}{!}{\includegraphics{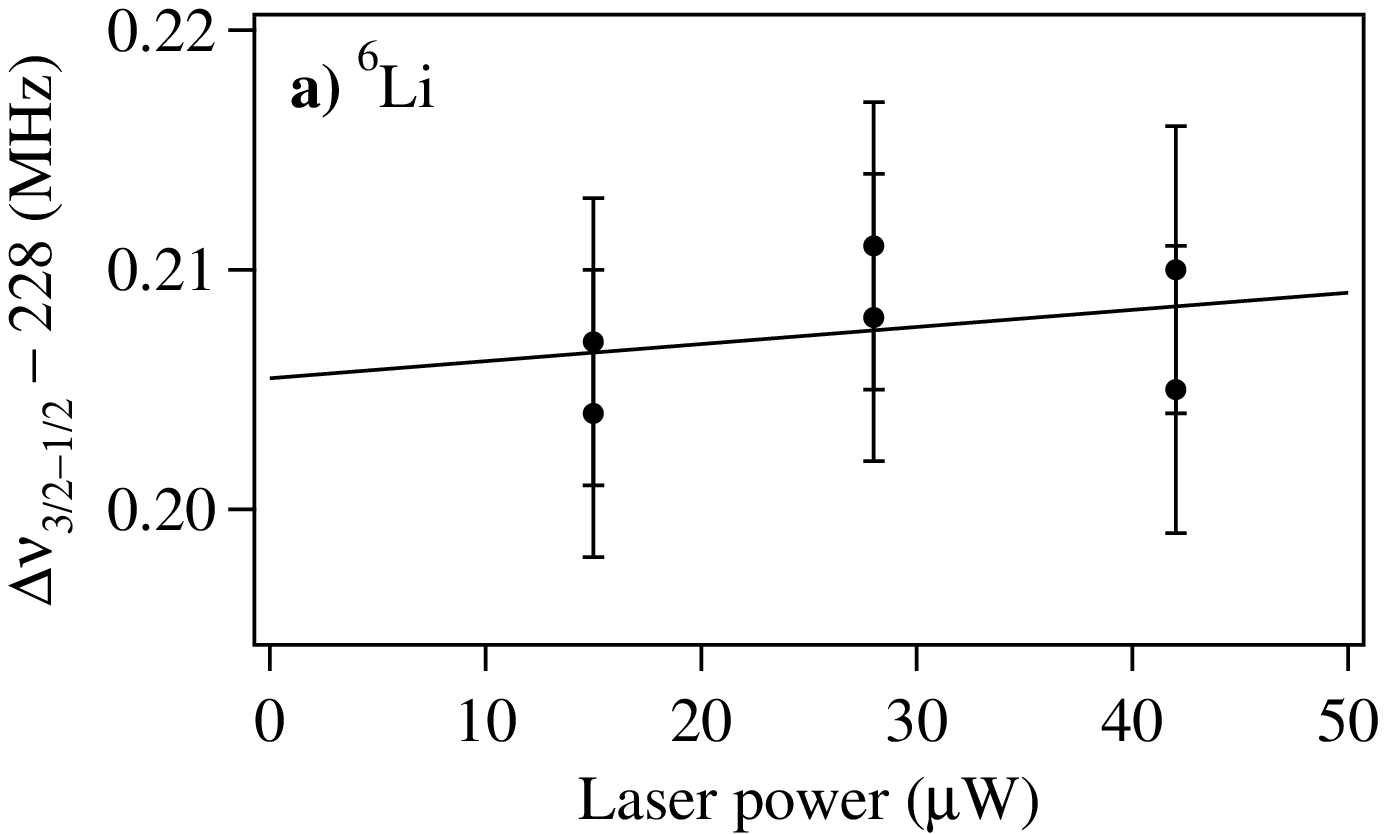}}
\resizebox{0.43\columnwidth}{!}{\includegraphics{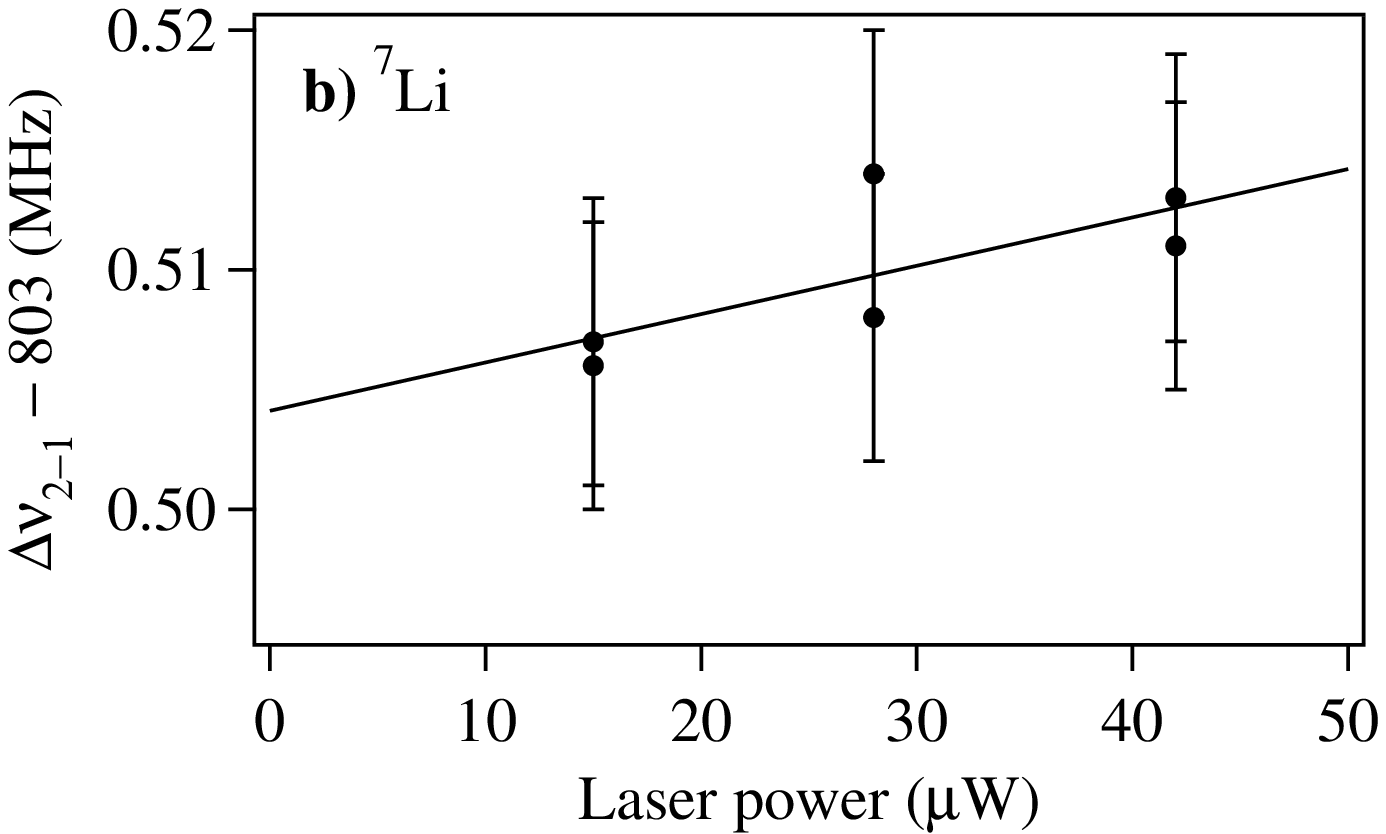}}
\caption{The measured hyperfine intervals in the ground
state are plotted against power to check for
intensity-dependent errors. The solid line is a linear fit
used to extrapolate to zero power.} \label{f2b}
\end{figure}

The important thing to note from the table is that all the
measurements lie within a few kHz of each other, even when
the laser power is increased by a factor of three. This
means that shifts due to radiation-pressure and
optical-pumping effects are less than 5 kHz, consistent
with our earlier estimate. Note that increasing the laser
power also increases the height of the peaks in the
spectrum, their linewidth, and the overall signal-to-noise
ratio. The consistency of the values shows that there are
no unknown systematic errors related to these parameters.
To see if there is any intensity-dependent trend in the
data, we have plotted the measured values versus power in
Fig.\ \ref{f2b}. The linear fit has a small slope, and the
extrapolated values of the intervals are
\begin{tabbing}
\hspace{0.75cm} $^{6}$Li, $2S_{1/2}$: \= $\Delta
\nu_{\frac{3}{2} - \frac{1}{2}} = 228.205(6)$ MHz.
\end{tabbing}
\begin{tabbing}
\hspace{0.75cm} $^{7}$Li, $2S_{1/2}$: \= \;$\Delta \nu_{2 -
1} = 803.504(6)$ MHz.
\end{tabbing}
The quoted error of 6 kHz is the total error obtained by
adding in quadrature the different sources of error in
Table \ref{t1}. All the values in the table lie within 6
kHz of the extrapolated value for $^{6}$Li, and 10 kHz for
$^{7}$Li. These values can be compared with the accurately
known values \cite{AIV77}: 228.205\,261\,1 MHz in $^{6}$Li
and 803.504\,086\,6 MHz in $^{7}$Li. Our values are
consistent within $1\sigma$, confirming that our error
estimate is reasonable.

We now turn to the measurement of hyperfine intervals in
the $2P_{1/2}$ state. Since there are two ground hyperfine
levels, the same interval can be measured with transitions
starting from either level. As before, the measurements
were done at three laser powers as listed in Table
\ref{t2}. The measurements at the different powers again
lie within a few kHz of each other. The measurements are
plotted against power to yield extrapolated values of the
intervals as
\begin{tabbing}
\hspace{0.75cm} $^{6}$Li, $2P_{1/2}$: \= $\Delta
\nu_{\frac{3}{2} - \frac{1}{2}} = \frac{3}{2}A$ \= $=
26.091(6)$ MHz.
\end{tabbing}
\begin{tabbing}
\hspace{0.75cm} $^{7}$Li, $2P_{1/2}$: \= \;$\Delta \nu_{2 -
1} = 2A$ \= $= 92.047(6)$ MHz.
\end{tabbing}
The maximum deviation of the values in the table from the
extrapolated values is only 8 kHz ($1.3 \sigma$). To the
extent that different transitions are susceptible to
different degrees of systematic error, this again confirms
that our error estimate is reasonable.

\begin{figure}
\hspace*{2cm}
\resizebox{0.43\columnwidth}{!}{\includegraphics{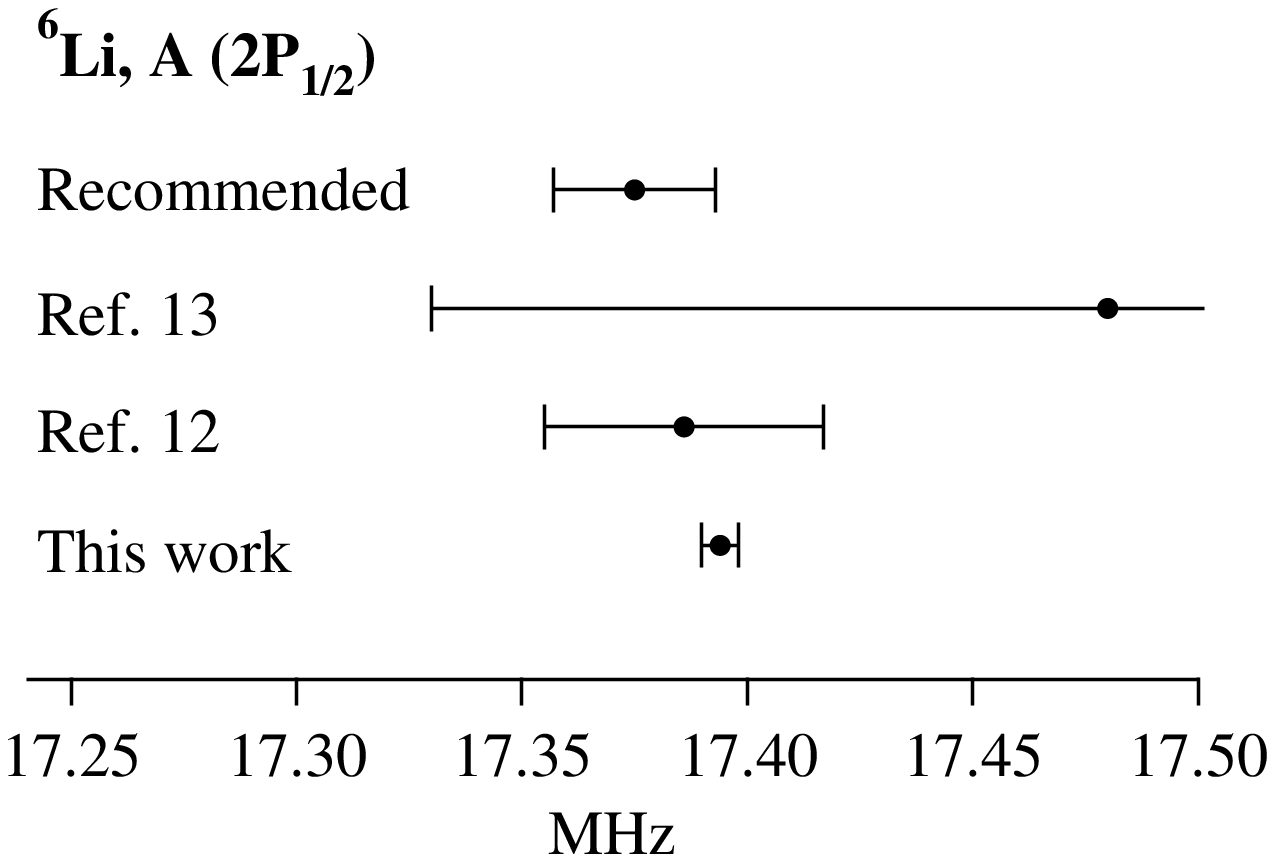}}
\resizebox{0.43\columnwidth}{!}{\includegraphics{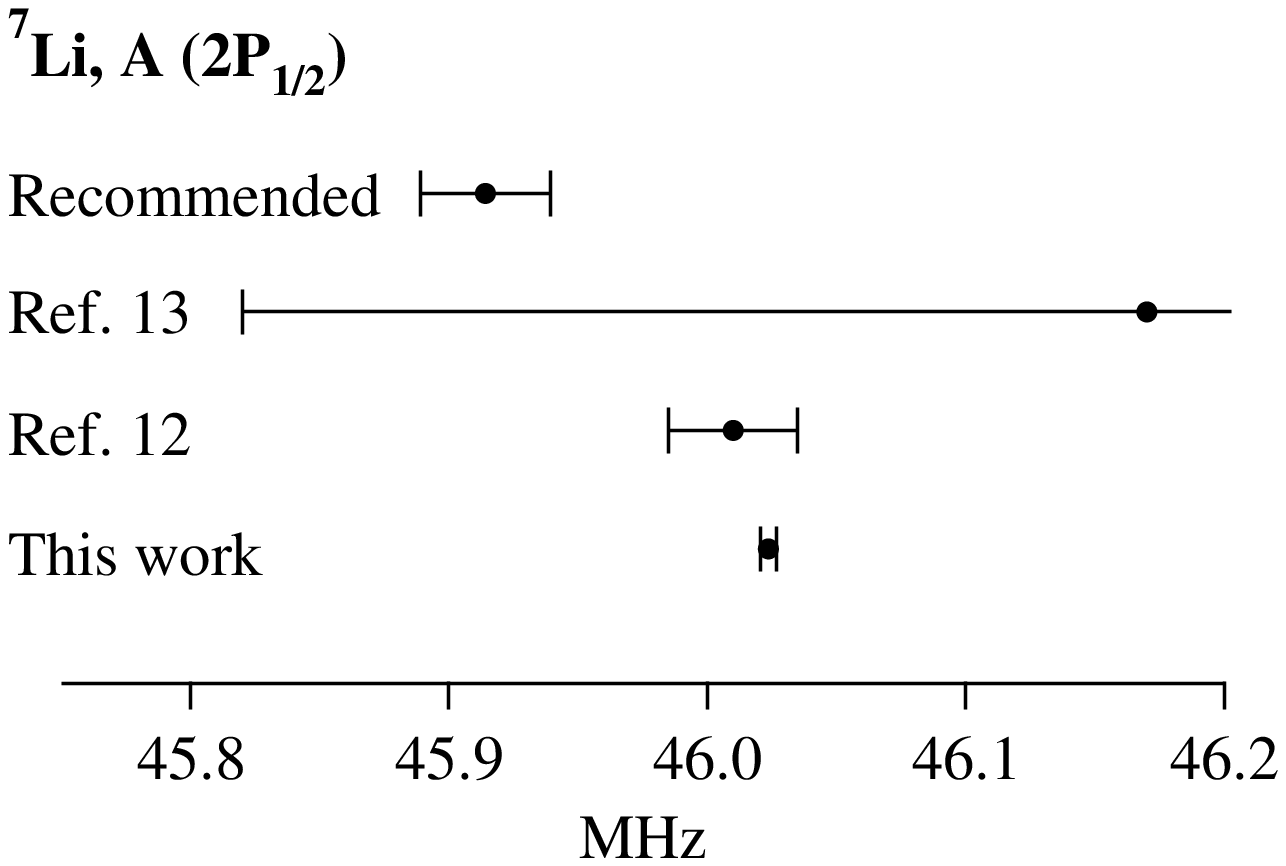}}
\caption{Comparison of the value of $A$ in the $2P_{1/2}$
state of $^{6,7}$Li obtained in this work to earlier
values. Also shown is the recommended value from Ref.\
\cite{AIV77}} \label{f2c}
\end{figure}

The value of the magnetic-dipole coupling constant $A$ in
the two isotopes is compared to previous values in Table
\ref{t4}. Also listed is the recommended value from the
review work of Arimondo, Inguscio, and Violino
\cite{AIV77}. The result in $^{6}$Li is completely
consistent with previous measurements, as seen from the
deviation plot in Fig.\ \ref{f2c}, but the accuracy is
improved by a factor of 5. In $^{7}$Li, our value is
slightly non-overlapping with the recommended value.
However, it is consistent with a recent measurement
\cite{WAC03}, which has the same accuracy as the
recommended value, and an earlier less precise measurement
\cite{RIT65}. Our accuracy is almost an order of magnitude
better. The last column in the table lists the result of a
recent calculation of $A$ in $^{7}$Li \cite{GFJ01}, which
agrees with our value at the 1.7\% level.

\section{Sodium}
Sodium has one stable isotope, $^{23}$Na, with $I=3/2$.
Both the $D$ lines are near 589 nm, and are accessed using
a cw ring-dye laser (Coherent 699-21) operating with
Rhodamine-6G dye. Na spectroscopy is done in a 100-mm long
vapor cell. The cell is heated to a temperature of
$75^\circ$C and has a magnetic shield around it.

\subsection{$D_1$ line}
The hyperfine levels in the $3P_{1/2}$ state are far apart
and the individual transitions in the $D_1$ line are well
resolved. Hence we have used the standard SAS technique for
spectroscopy to lock the laser and the AOM. As in the case
of Li, intensity-dependent errors are checked by repeating
the measurements at three laser powers. The average value
of the interval is
\begin{tabbing}
\hspace{0.75cm} $^{23}$Na, $3P_{1/2}$: \= $\Delta \nu_{2 -
1} = 2A$ \= $= 188.697(14)$ MHz.
\end{tabbing}

\begin{figure}
\hspace*{2cm}
\resizebox{0.43\columnwidth}{!}{\includegraphics{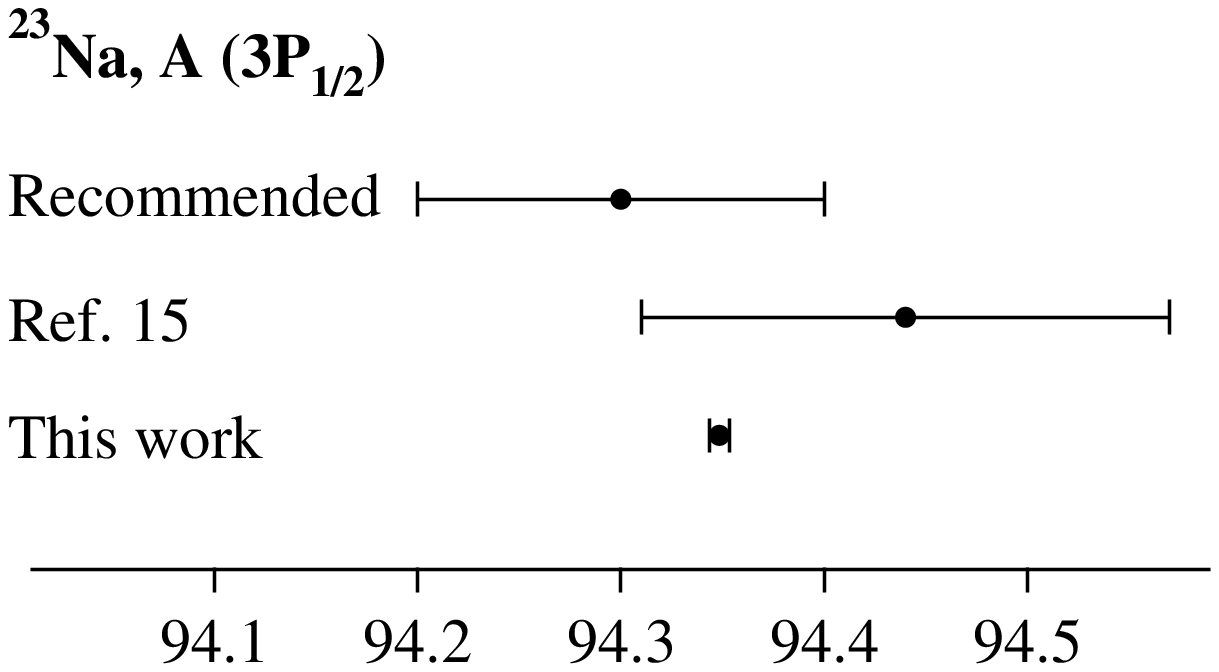}}
\caption{Comparison of the value of $A$ in the $3P_{1/2}$
state of $^{23}$Na obtained in this work to earlier
values.} \label{f3a}
\end{figure}

The value of the hyperfine constant is compared to the
recommended value and another published result \cite{WIL94}
in Fig.\ \ref{f3a}. Our value is consistent with these
results but the accuracy is improved by a factor of 20.

\subsection{$D_2$ line}
As mentioned earlier, the hyperfine levels in the
$3P_{3/2}$ state of $^{23}$Na are too close for the
individual transitions in the $D_2$ line to be resolved.
Hence we have used the technique of CCS to resolve these
transitions. The results of our measurements have been
published previously \cite{DPW06}. The average values of
the intervals are:
\begin{tabbing}
\hspace{0.75cm} $^{23}$Na, $3P_{3/2}$:
\= $\Delta \nu_{3 - 2}$ \= $= 3A + B = 58.310(11)$ MHz, \\
\> $\Delta \nu_{2 - 1}$ \> $= 2A - B = 34.339(11)$ MHz.
\end{tabbing}

\begin{figure}
\hspace*{2cm}
\resizebox{0.43\columnwidth}{!}{\includegraphics{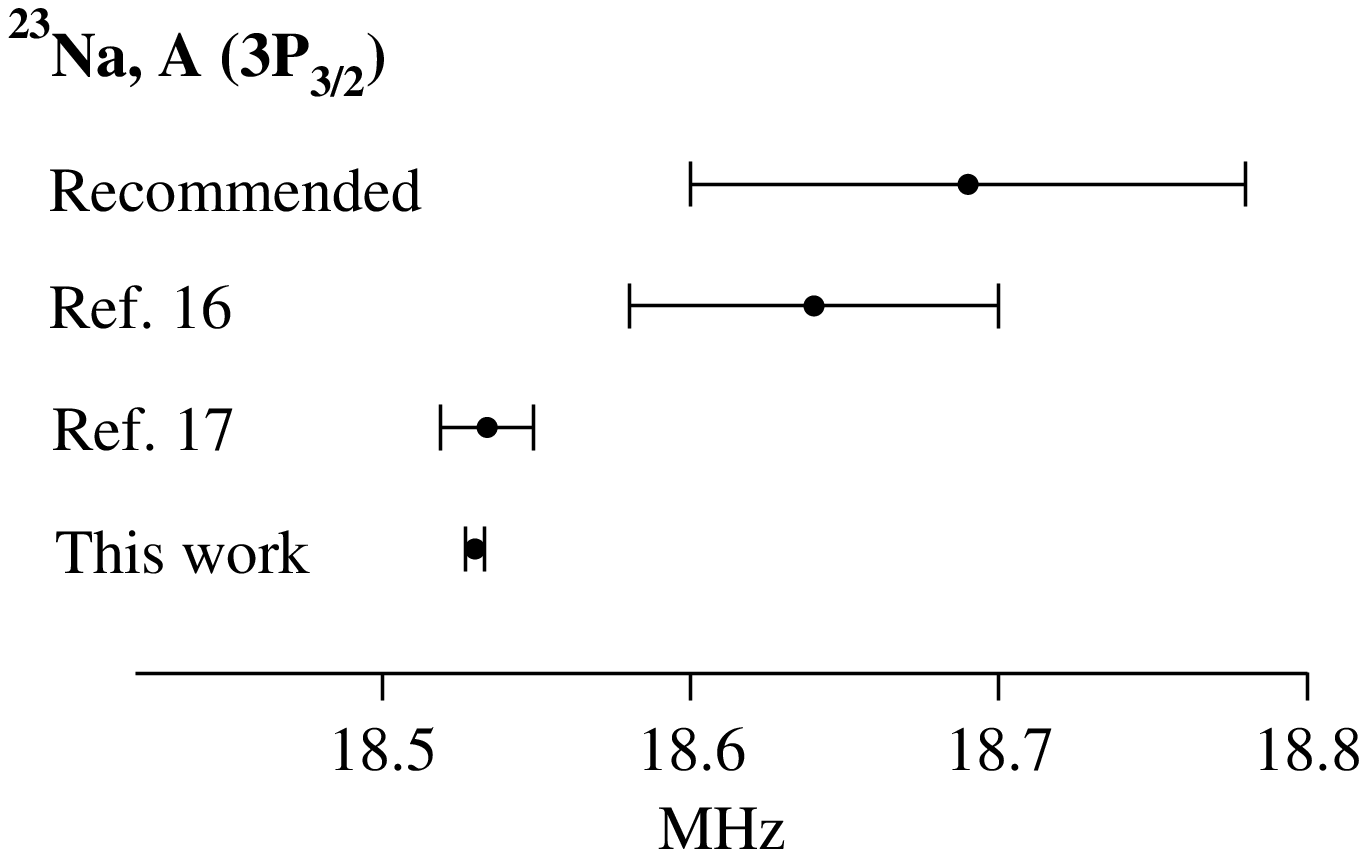}}
\resizebox{0.43\columnwidth}{!}{\includegraphics{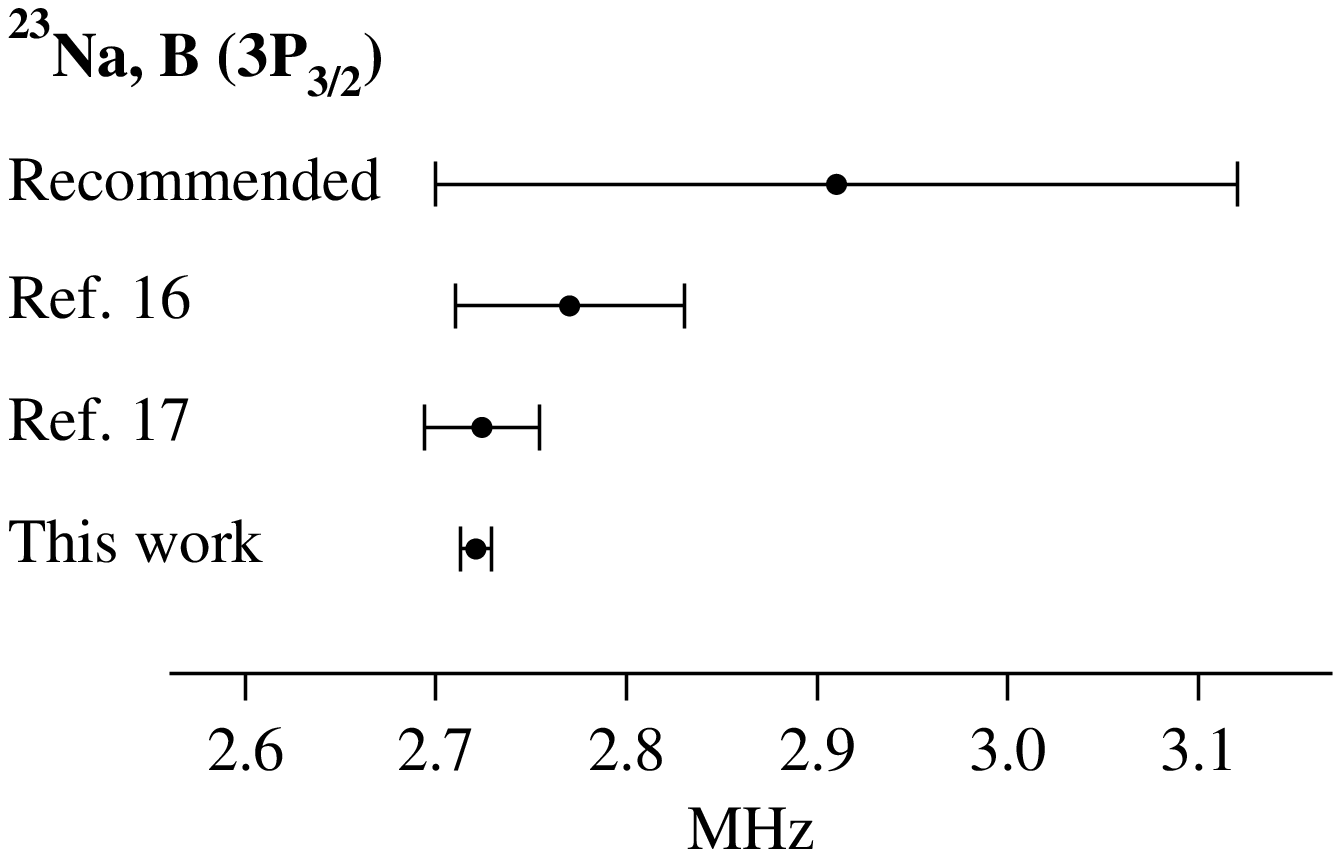}}
\caption{Comparison of the values of $A$ and $B$ in the
$3P_{3/2}$ state of $^{23}$Na obtained in this work to
earlier values.} \label{f3b}
\end{figure}

In Fig.\ \ref{f3b}, we compare our values of the hyperfine
constants to the recommended value and two subsequent
measurements, one using time-resolved hyperfine
quantum-beat spectroscopy \cite{KKG77} and the second using
polarization quantum-beat spectroscopy \cite{YSH93}. Our
values are consistent with the earlier values but the
accuracy is improved by a factor of 5.

\section{Potassium}
Potassium has two stable isotopes: $^{39}$K with $I=3/2$
and natural abundance of 93.3\%, and $^{41}$K with $I=3/2$
and natural abundance of 6.7\%. The $D_1$ line is at 770 nm
and the $D_2$ line is at 767 nm. It is difficult to obtain
reliable laser diodes at these wavelengths, hence our
experiments were done sometimes with a frequency-stabilized
diode and sometimes with a ring-cavity Ti:sapphire laser
(Coherent 899-21) similar to the ring dye laser used for
the sodium experiments. Potassium spectroscopy was done in
a 50-mm long vapor cell. The cell was heated to a
temperature of $75^\circ$C and had a magnetic shield around
it. Due to the low natural abundance of $^{41}$K, it is
difficult to get a good absorption signal for this isotope,
hence we have measured hyperfine structure only in
$^{39}$K.

\subsection{$D_1$ line}
The hyperfine levels in the $4P_{1/2}$ state are separated
by about 58 MHz compared to the natural linewidth of 6 MHz.
Thus the individual transitions in the $D_1$ line are
fairly well resolved. In a first set of experiments, we
measured the interval using normal SAS to lock the laser
and the AOM. Note that the crossover resonance is only 29
MHz away, which is not very far given that the
experimentally obtained linewidth in SAS is around 10 MHz.
The result for the interval from these measurements was
$57.723(15)$ MHz. The slightly large error is due to the
nearby crossover resonance which causes some peak pulling.

In an earlier measurement from our laboratory
\cite{BDN04b}, we had measured the absolute frequencies of
individual transitions in the $D_1$ line and hence
determined the hyperfine interval indirectly. The direct
measurement of the interval was consistent with this
earlier result. In addition, it agreed with the recommended
value in Ref.\ \cite{AIV77} (from a measurement by Buck and
Rabi \cite{BUR59} using the atomic beam magnetic resonance
technique). However, there has since been a report of
absolute frequency measurements on the same line using a
femtosecond comb \cite{FTL06}. The value of the interval
disagrees from our value (by $25 \sigma$ combined) and the
recommended value (by $3.5 \sigma$ combined), too large to
be accounted for by any known sources of error. We have
therefore repeated the measurement of the interval but this
time using the CCS technique. As mentioned earlier, this
technique does not produce crossover resonances and the
measured interval is insensitive to detuning of the control
laser.

\begin{figure}
\hspace*{2cm}
\resizebox{0.45\columnwidth}{!}{\includegraphics{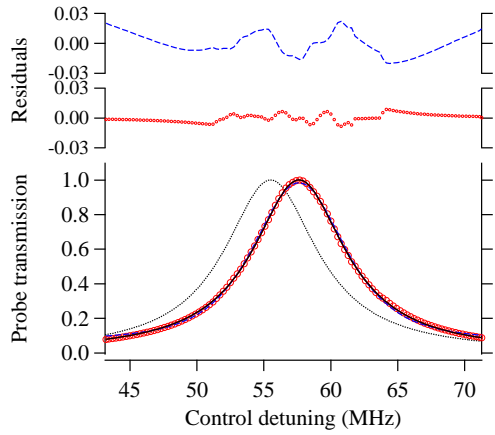}}
\caption{(Color online) $D_1$ spectrum of $^{39}$K. The
open circles show the probe transmission spectrum
(normalized) using the CCS technique obtained with the
probe beam locked to the $F=2 \rightarrow 2$ transition and
the control beam scanning across the $F=2 \rightarrow 1$
transition. The solid curves are fits with residuals shown
on top. The lower residuals are from a Lorentzian fit while
the upper ones are from a density-matrix analysis. The fits
yield a value of 57.713(42) MHz for the interval. The
dashed curve is the spectrum that would be obtained if the
interval were 55.500(84) MHz as measured by Falke {\it et
al.} \cite{FTL06}.} \label{f4a}
\end{figure}

In Fig.\ \ref{f4a}, we show a typical probe transmission
spectrum in $^{39}$K obtained with the probe locked to the
$F=2 \rightarrow 2$ transition and the control scanning
across the $F=2 \rightarrow 1$ transition. The scanning is
done by varying the frequency of the AOM in the path of the
control, so that the scan axis is properly calibrated and
its linearity is guaranteed by the linearity of the
voltage-controlled oscillator (vco) driving the AOM. Note
that direct scanning of the laser using intracavity
elements usually results in a nonlinear scan. The AOM is
double passed to ensure that the direction of the control
beam remains unchanged while the AOM is scanned, and the
output beam intensity is stabilized by feedback control of
the rf power exciting the AOM.

The open circles in the figure represent the measured
spectrum while the solid curves are fits. We did two kinds
of fitting to the spectrum. The first fit is to a
Lorentzian lineshape with the peak center and linewidth as
fit parameters. The second fit is to a density-matrix
calculation of the effective three-level system taking into
account the thermal velocity distribution in hot vapor, as
described in our earlier work \cite{DAN05}. The hyperfine
interval (which determines the control detuning at the
peak) and the control Rabi frequency are free fit
parameters. The residuals from the two fits are also shown
in the figure, with the Lorentzian fit yielding slightly
smaller residuals. However, the location of the peak center
from both fits is the same. After fitting to several
independent spectra (to take care of drifts in the vco
driving the AOM), we obtain a value of 57.713(42) MHz for
the interval, close to our earlier measurement and the one
by Buck and Rabi. The dashed curve shows what the spectrum
would look like if the interval were 55.500 MHz as measured
by Falke {\it et al.}

Although the spectrum obtained by scanning the AOM yields
the interval, the locking technique has higher accuracy
because the AOM frequency is measured while it is locked,
and is not susceptible to drifts of the vco. The new result
for the interval by locking the AOM is
\begin{tabbing}
\hspace{0.75cm} $^{39}$K, $4P_{1/2}$: \= $\Delta \nu_{2 -
1} = 2A$ \= $= 57.696(10)$ MHz,
\end{tabbing}
which is again consistent with the above measurements and
our earlier published value.

\begin{figure}
\hspace*{2cm}
\resizebox{0.5\columnwidth}{!}{\includegraphics{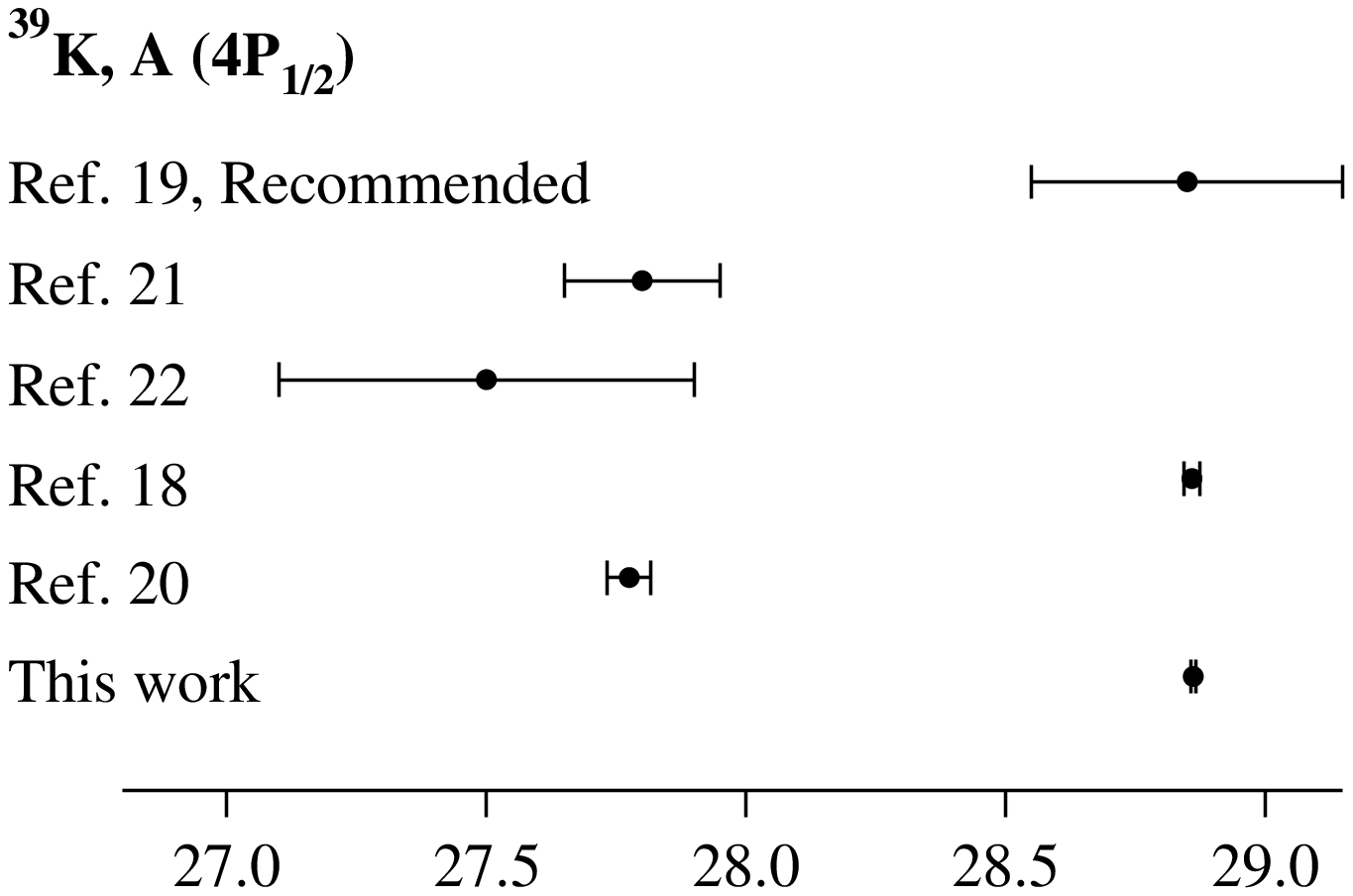}}
\caption{Comparison of the value of $A$ in the $4P_{1/2}$
state of $^{39}$K obtained in this work to earlier values.}
\label{f4b}
\end{figure}

In  Fig.\ \ref{f4b}, we compare this value of $A$ with
other published values. As can be seen, there are two sets
of values that are in complete disagreement with each
other. The recommended value and our two results are
clustered on one side, while the recent result of Falke
{\it et al}.\ \cite{FTL06} and two older measurements
(which are both from the same group \cite{BDV81,TGB82}) are
clustered together. The experiment of Falke {\it et al}.\
\cite{FTL06} has also been done with great care and takes
into account many potential sources of error, hence their
measurement seems to be reliable within the stated
accuracy. We do not have a satisfactory explanation for
such a large discrepancy.

\subsection{$D_2$ line}
As mentioned before, the hyperfine levels in the $4P_{3/2}$
state of $^{39}$K are too close to be resolved completely.
Hence we have used our technique of CCS to get a
partially-resolved spectrum. However, the peaks are still
so close that the lock point will get pulled by the nearby
transition. Therefore, as described above for the $D_1$
spectrum, we measure the intervals by scanning the AOM in
the path of the control beam. The AOM is again
double-passed and the output beam is intensity stabilized.

A typical probe transmission spectrum of the $D_2$ line
with the probe locked on the $F=2 \rightarrow F'=3$
transition is shown in Fig.\ \ref{f4b}. We saw above in the
case of the $D_1$ line that the lineshape is described
quite well by a Lorentzian curve, hence we extract the
locations of the individual hyperfine transitions by doing
a multipeak fit to the spectrum. As seen from the figure,
the featureless residuals are less than 1\% of the peak.
The fit linewidth is 8.4 MHz, or only 45\% larger than the
natural linewidth of 5.8 MHz. After fitting to 25 spectra
we obtain the average values of the intervals as
\begin{tabbing}
\hspace{0.75cm} $^{39}$K,  $4P_{3/2}$:
\= $\Delta \nu_{3 - 2}$ \= $= 3A + B = 21.107(90)$ MHz, \\
\> $\Delta \nu_{2 - 1}$ \> $= 2A - B = 9.280(70)$ MHz.
\end{tabbing}
The slightly larger errors are due to the drift of the vco
from shot to shot.

\begin{figure}
\hspace*{2cm}
\resizebox{0.45\columnwidth}{!}{\includegraphics{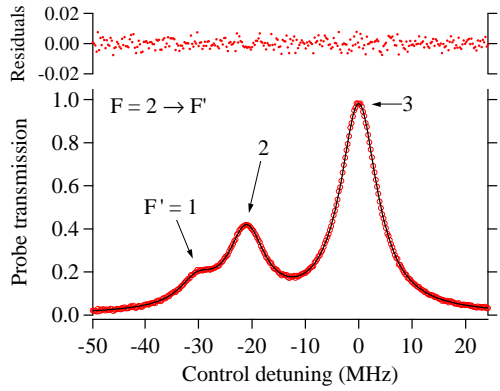}}
\caption{(Color online) $D_2$ spectrum (normalized) of
$^{39}$K without crossover resonances using the
coherent-control technique. The control detuning is
measured from the $F'=3$ peak. The open circles are the
observed spectrum, while the solid curve is a three peak
Lorentzian fit which yields the hyperfine intervals.}
\label{f4c}
\end{figure}

The difficulty in measuring hyperfine structure in this
state is evidenced by the fact that there has only been one
other measurement in recent times \cite{FTL06}. Our values
of the hyperfine constants $A$ and $B$ are compared to this
measurement and the recommended values in Fig.\ \ref{f4c}.
The three sets are in complete agreement with each other,
with our values having the smallest uncertainty. Note that
the recent values in Ref.\ \cite{FTL06} are from the same
work which differed so significantly for the $D_1$ line.

\section{Rubidium}
Rubidium has two stable isotopes: $^{85}$Rb with $I=5/2$
and natural abundance of 72.2\%, and $^{87}$Rb with $I=3/2$
and natural abundance of 27.8\%. The $D_1$ line at 795 nm
and the $D_2$ line at 780 nm are both accessed using
frequency-stabilized diode lasers. Rb spectroscopy is done
in a 50-mm long vapor cell containing both isotopes. The
cell is at room temperature and has a magnetic shield
around it.

\subsection{$D_1$ line}
The hyperfine levels in the $5P_{1/2}$ state are far apart
and the individual transitions in the $D_1$ line are well
resolved. Therefore we have used the standard SAS technique
for spectroscopy. The results of our measurements have been
published before \cite{DAN06a}, with the average values in
the two isotopes as:
\begin{tabbing}
\hspace{0.75cm} $^{85}$Rb, $5P_{1/2}$: \= $\Delta \nu_{3 -
2} = 3A$ \= $= 361.936(14)$ MHz.
\end{tabbing}
\begin{tabbing}
\hspace{0.75cm} $^{87}$Rb, $5P_{1/2}$: \= $\Delta \nu_{2 -
1} = 2A$ \= $= 812.238(14)$ MHz.
\end{tabbing}

\begin{figure}
\hspace*{2cm}
\resizebox{0.43\columnwidth}{!}{\includegraphics{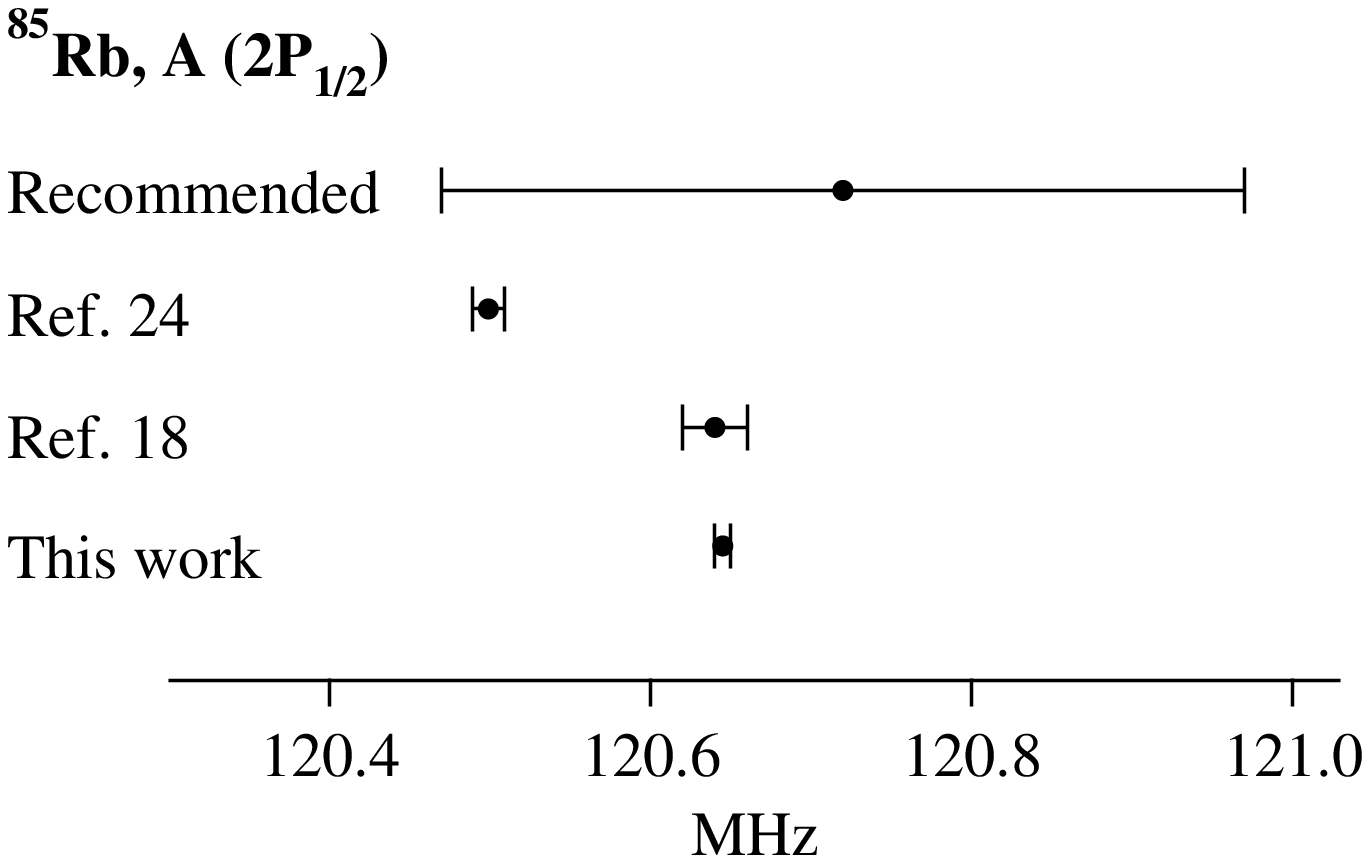}}
\resizebox{0.43\columnwidth}{!}{\includegraphics{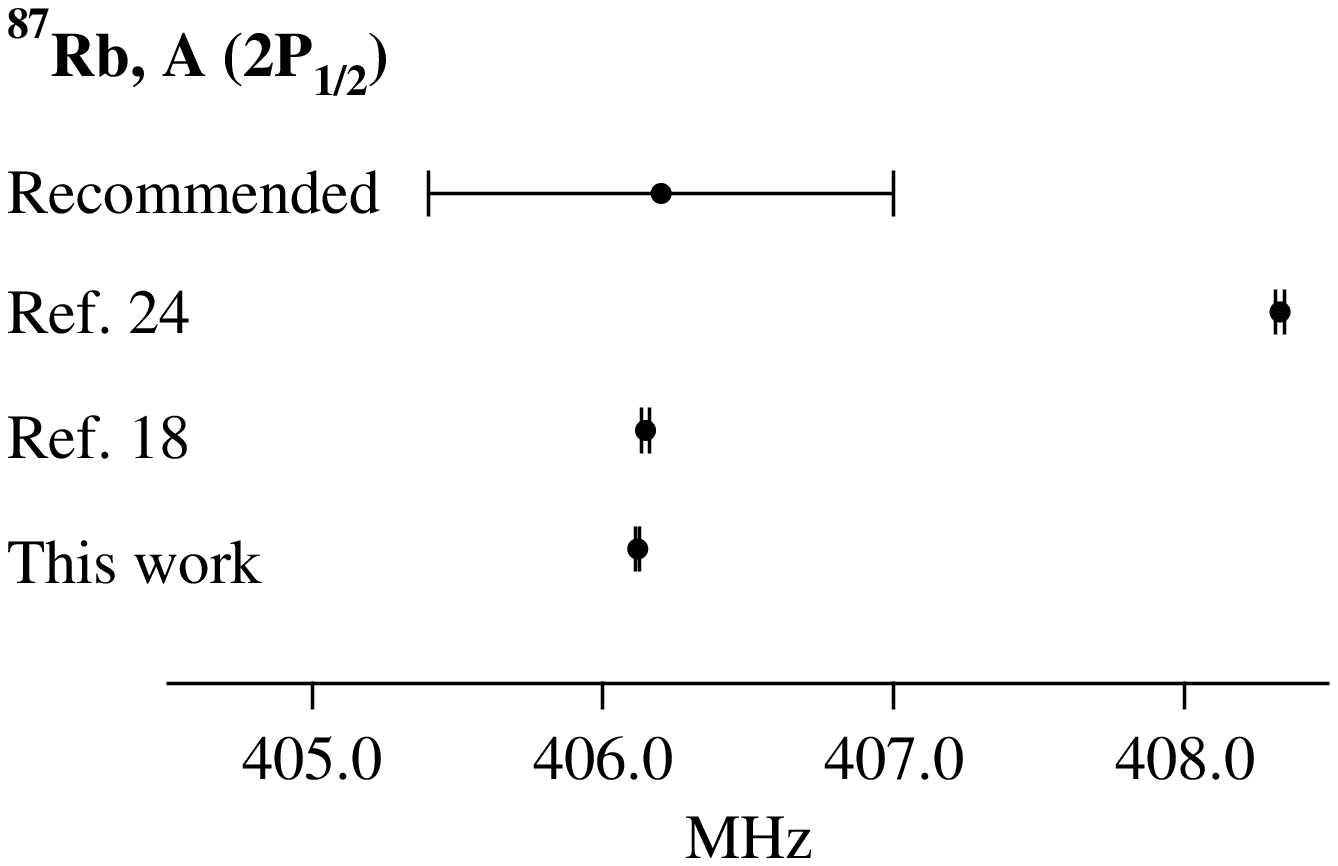}}
\caption{Comparison of the value of $A$ in the $5P_{1/2}$
state of $^{85,87}$Rb obtained in this work to earlier
values.} \label{f5a}
\end{figure}

These values are compared to earlier values in Fig.\
\ref{f5a}. One of the reasons for our interest in Rb was
that the values from Ref.\ \cite{BGR91} differed
significantly from an earlier measurement from our
laboratory \cite{BDN04b}. In both these cases, the absolute
frequencies of the individual transitions were measured and
the hyperfine interval determined from their difference. By
contrast, in the present work we measure the interval
directly. As can be seen from the figure, the present value
agrees with our earlier indirect measurement, but disagrees
with the one from Ref.\ \cite{BGR91}. Note that the
value for $^{87}$Rb from Ref.\ \cite{BGR91} also disagrees
with the recommended value.

\subsection{$D_2$ line}
The hyperfine levels in the $5P_{3/2}$ state are far apart
and the individual transitions in the $D_2$ line are well
resolved. Therefore we have used the standard SAS technique
for spectroscopy. As usual, we check for
intensity-dependent errors by repeating the measurements at
several powers. We have earlier published results in
$^{85}$Rb using a similar technique to measure the
intervals \cite{RKN03}. Although those results were the
most accurate then, our current work improves on that
precision significantly. In addition, we report results in
$^{87}$Rb for the first time. The average values for the
intervals are:
\begin{tabbing}
\hspace{0.75cm} $^{85}$Rb, $5P_{3/2}$:
\= $\Delta \nu_{3 - 2}$ \= $= 3A - \frac{9}{20}B$ \= $= 63.424(6)$ MHz. \kill
\hspace{0.75cm} $^{85}$Rb, $5P_{3/2}$:
\> $\Delta \nu_{4 - 3}$ \> $= 4A + \frac{4}{5}B$ \> $= 120.966(8)$ MHz, \\
\> $\Delta \nu_{3 - 2}$ \> $= 3A - \frac{9}{20}B$ \> $= 63.424(6)$ MHz, \\
\> $\Delta \nu_{2 - 1}$ \> $= 2A - \frac{4}{5}B$ \> $= 29.268(7)$ MHz.
\end{tabbing}

\begin{tabbing}
\hspace{0.75cm} $^{87}$Rb, $5P_{3/2}$:
\= $\Delta \nu_{3 - 2}$ \= $= 3A + B$ \= $= 266.657(8)$ MHz, \\
\> $\Delta \nu_{2 - 1}$ \> $= 2A - B$ \> $= 156.943(8)$ MHz, \\
\> $\Delta \nu_{1 - 0}$ \> $= A - B $ \> $= 72.223(9)$ MHz.
\end{tabbing}

\begin{figure}
\hspace*{2cm}
\resizebox{0.43\columnwidth}{!}{\includegraphics{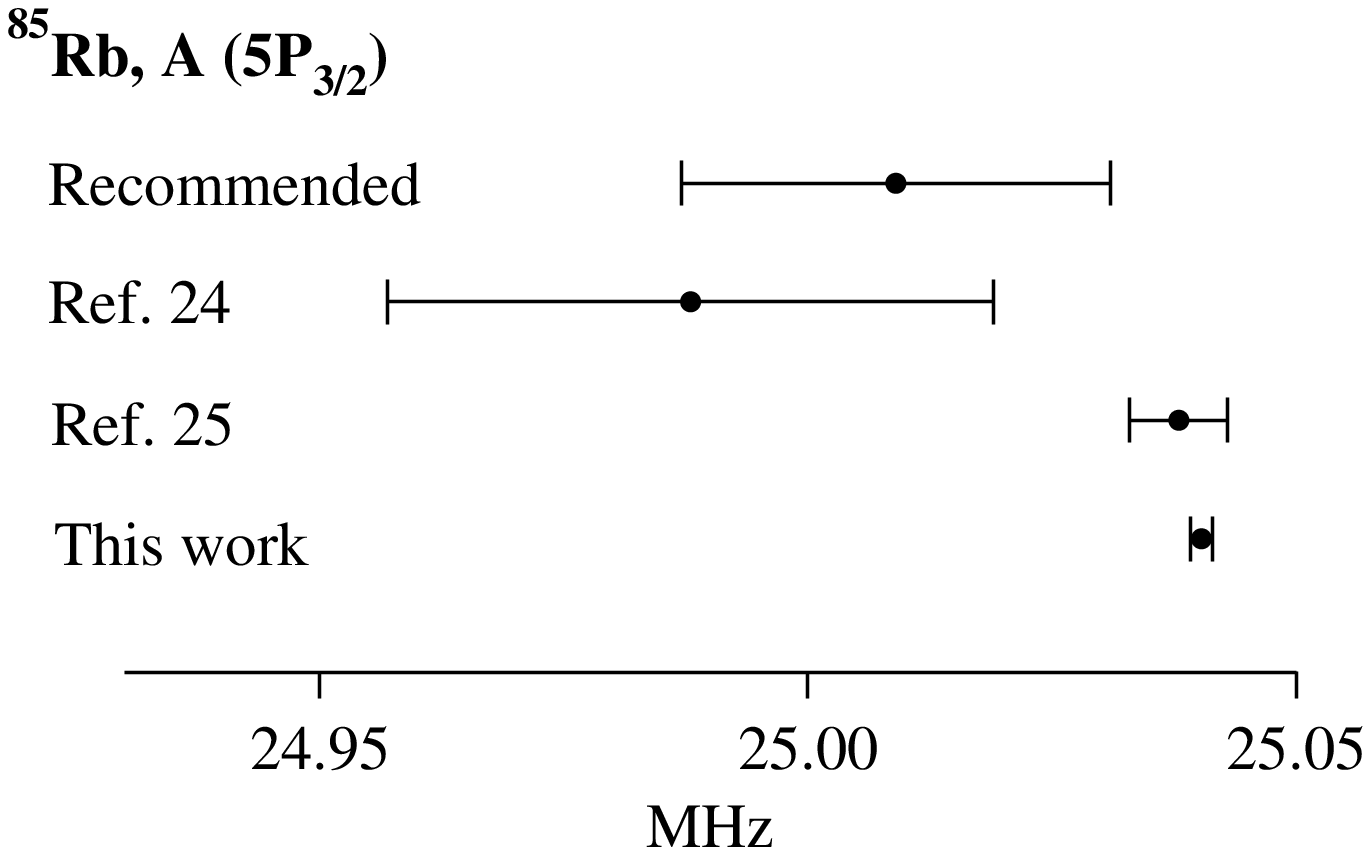}}
\resizebox{0.43\columnwidth}{!}{\includegraphics{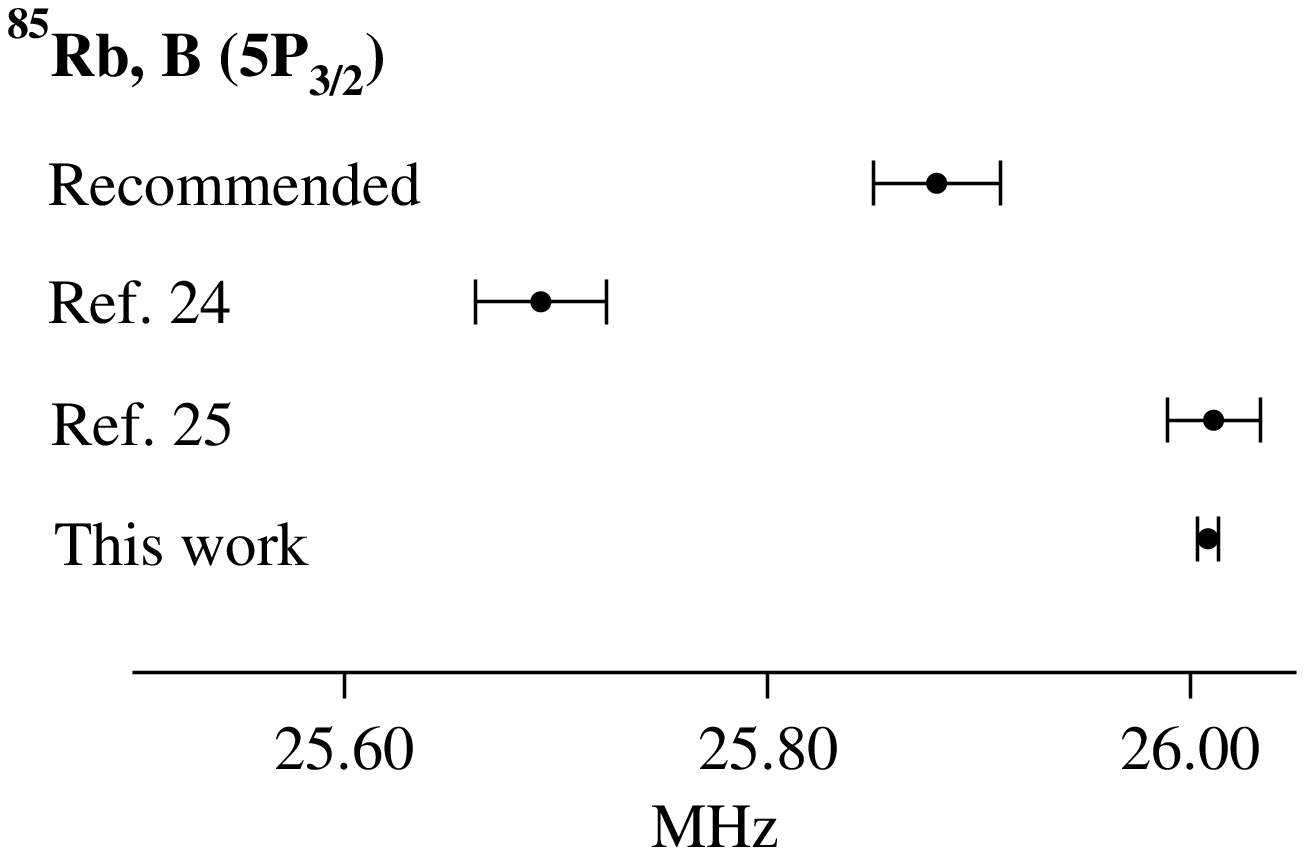}}
\hspace*{2cm}
\resizebox{0.43\columnwidth}{!}{\includegraphics{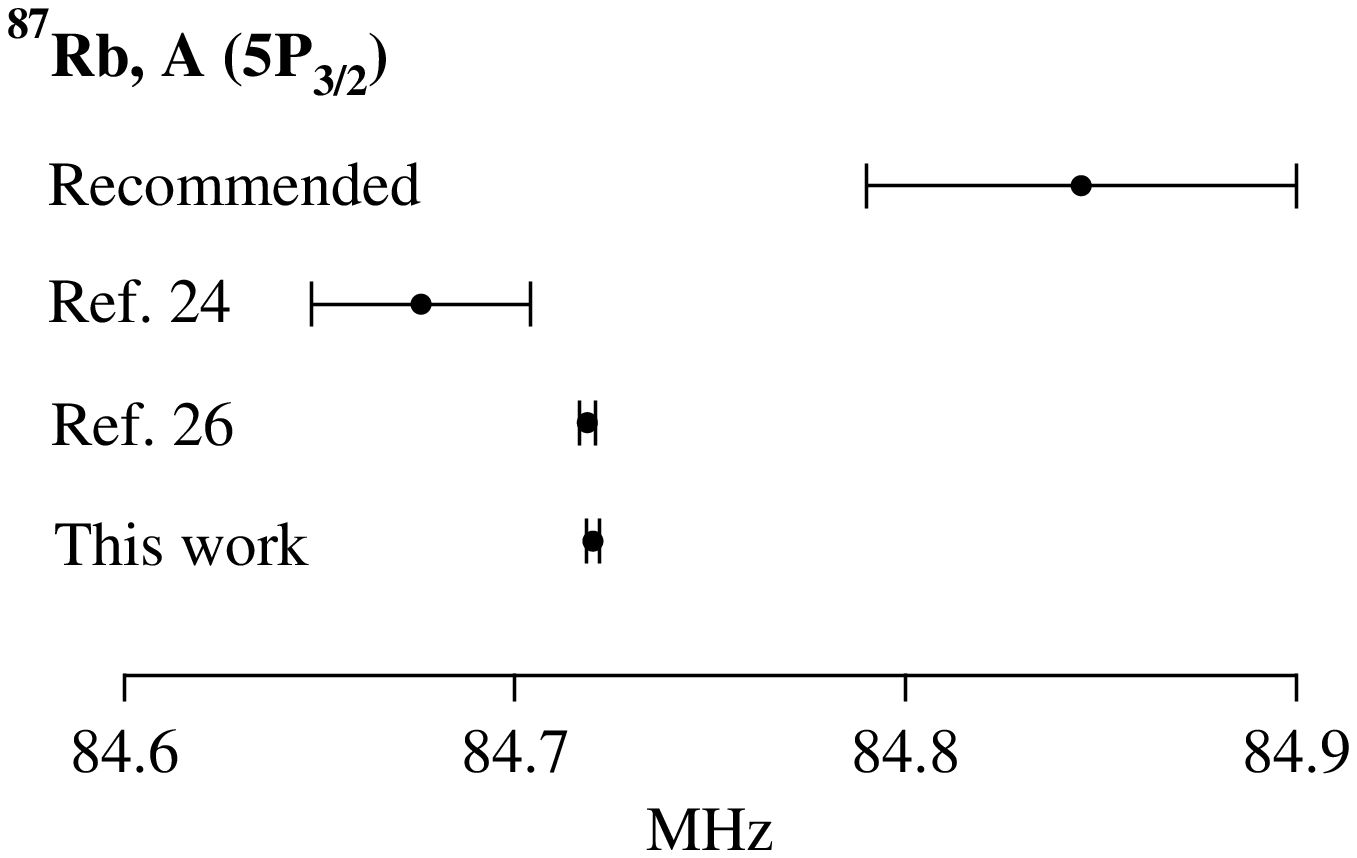}}
\resizebox{0.43\columnwidth}{!}{\includegraphics{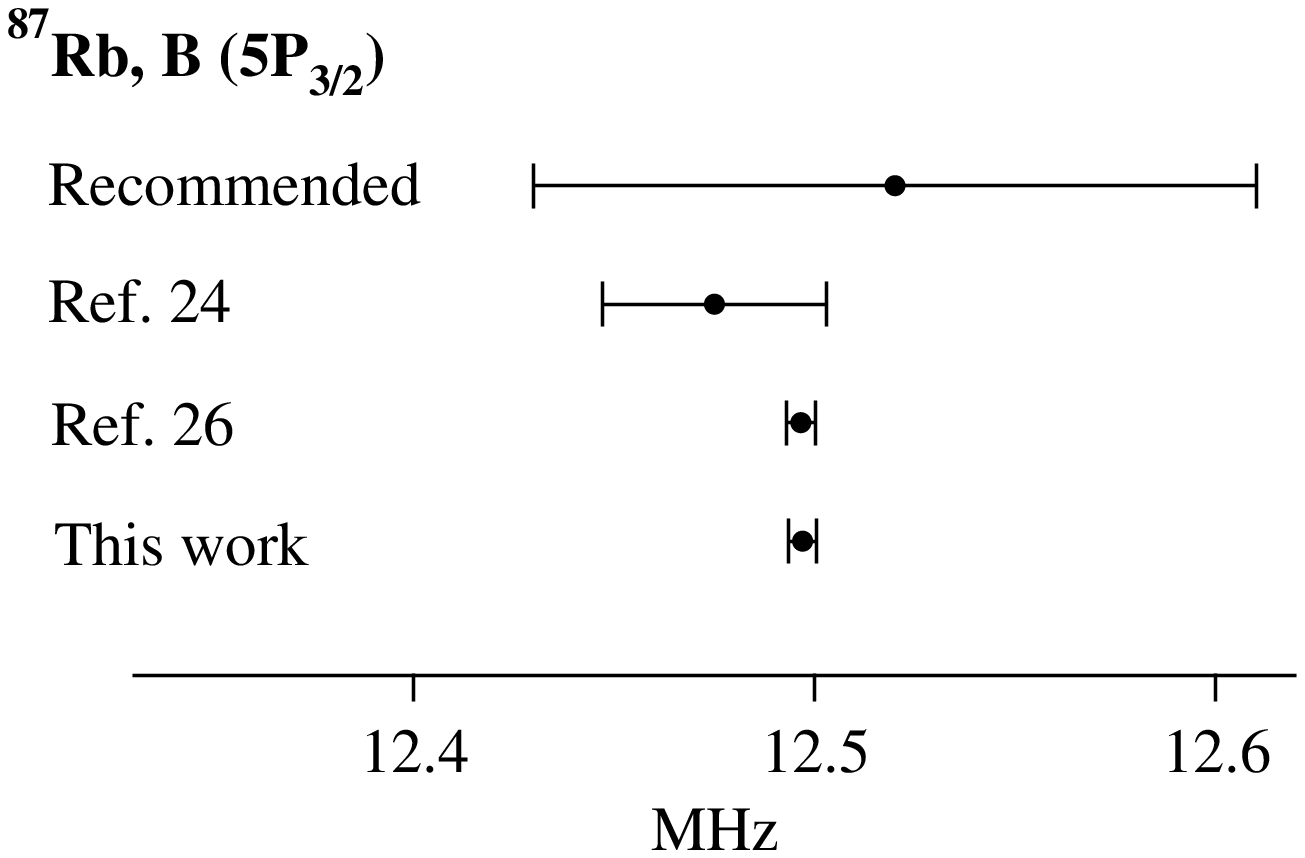}}
\caption{Comparison of the values of $A$ and $B$ in the
$5P_{3/2}$ state of $^{85,87}$Rb obtained in this work to
earlier values.} \label{f5b}
\end{figure}

The hyperfine constants are compared to previous values in
Fig.\ \ref{f5b}. In $^{85}$Rb, these values are consistent
with our earlier measurement and the recommended values.
For $B$, there is a slight discrepancy from the value in
Ref.\ \cite{BGR91}, but this is the same work where there
was a large discrepancy in the $D_1$ line as well. For
$^{87}$Rb, the three recent high-precision measurements are
in good agreement with each other. The values in Ref.\
\cite{YSJ96} have roughly the same accuracy as ours, and
are obtained from absolute frequency measurements of
various transitions in the $D_2$ line. Indeed, their
measurement of the $D_2$ frequency with a precision of $1.5
\times 10^{-11}$ has enabled us to use this line as a
frequency reference along with a ring-cavity resonator for
absolute optical frequency measurements of other
transitions \cite{BDN03,DBB06}.

\section{Cesium}
Cs has just one stable isotope, $^{133}$Cs with $I=7/2$.
The $D_1$ line at 895 nm and the $D_2$ line at 852 nm are
both accessed using frequency-stabilized diode lasers. Cs
spectroscopy is done in a 50-mm long vapor cell. The cell
is at room temperature and has a magnetic shield around it.

\subsection{$D_1$ line}
The hyperfine levels in the $6P_{1/2}$ state are more than
1 GHz apart and the individual transitions in the $D_1$
line are well resolved. Therefore we have used the standard
SAS technique to lock the laser and the AOM. The hyperfine
interval is actually larger than the Doppler width, hence
the spectrum has no crossover resonances. The results have
been published previously \cite{DAN06b}, with the average
value of the hyperfine interval as
\begin{tabbing}
\hspace{0.75cm} $^{133}$Cs, $6P_{1/2}$: \= $\Delta \nu_{4 -
3} = 4A$ \= $= 1167.654(6)$ MHz.
\end{tabbing}

\begin{figure}
\hspace*{2cm}
(a)\resizebox{0.43\columnwidth}{!}{\includegraphics{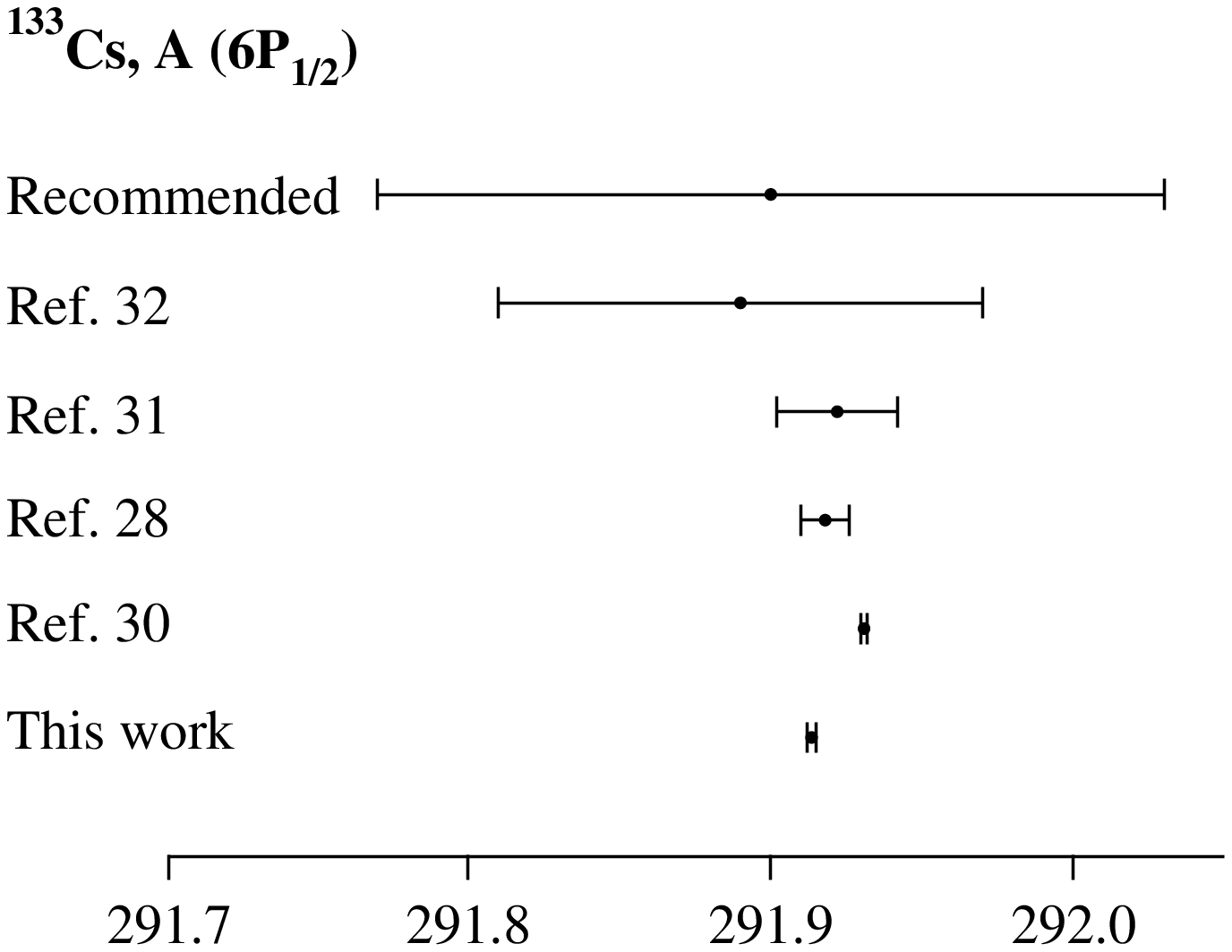}} \\
\hspace*{2cm}
(b)\resizebox{0.43\columnwidth}{!}{\includegraphics{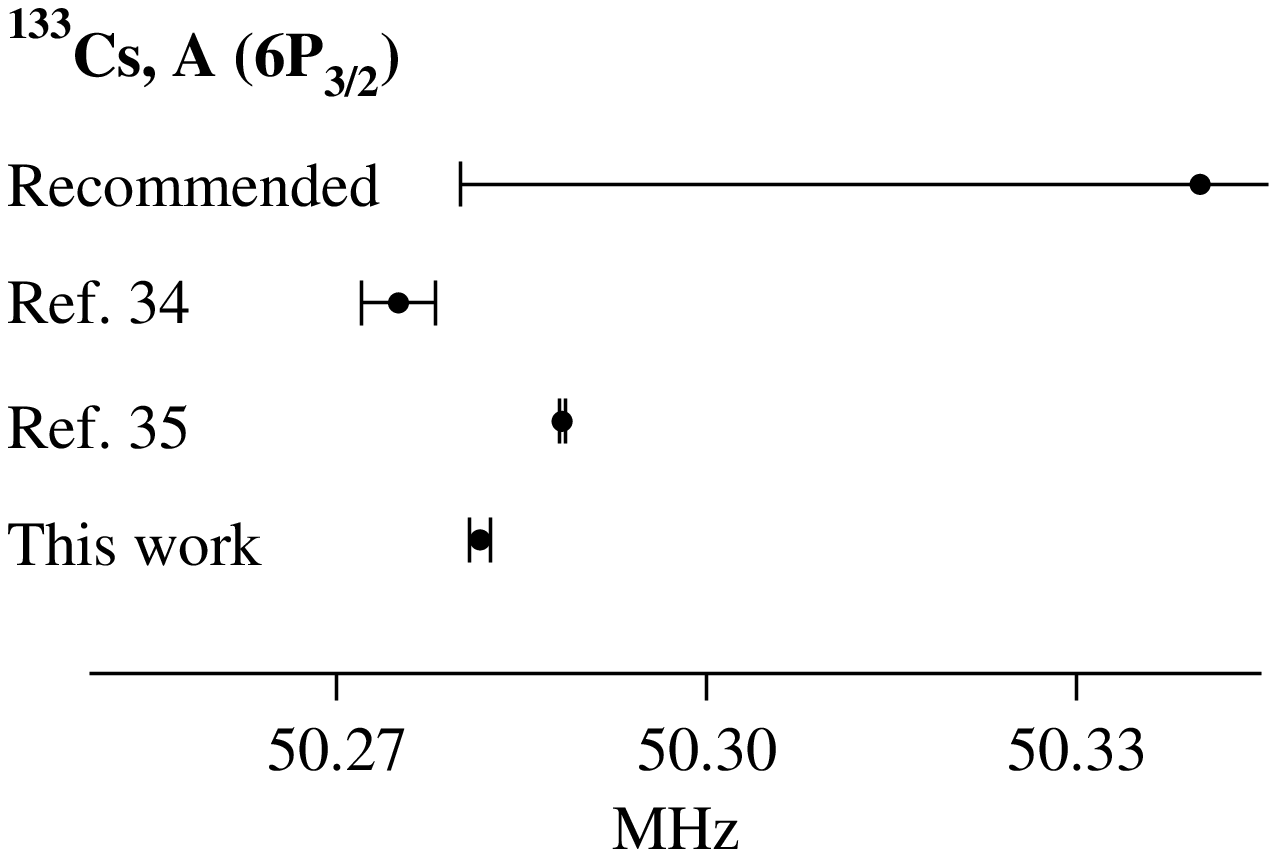}}
\resizebox{0.43\columnwidth}{!}{\includegraphics{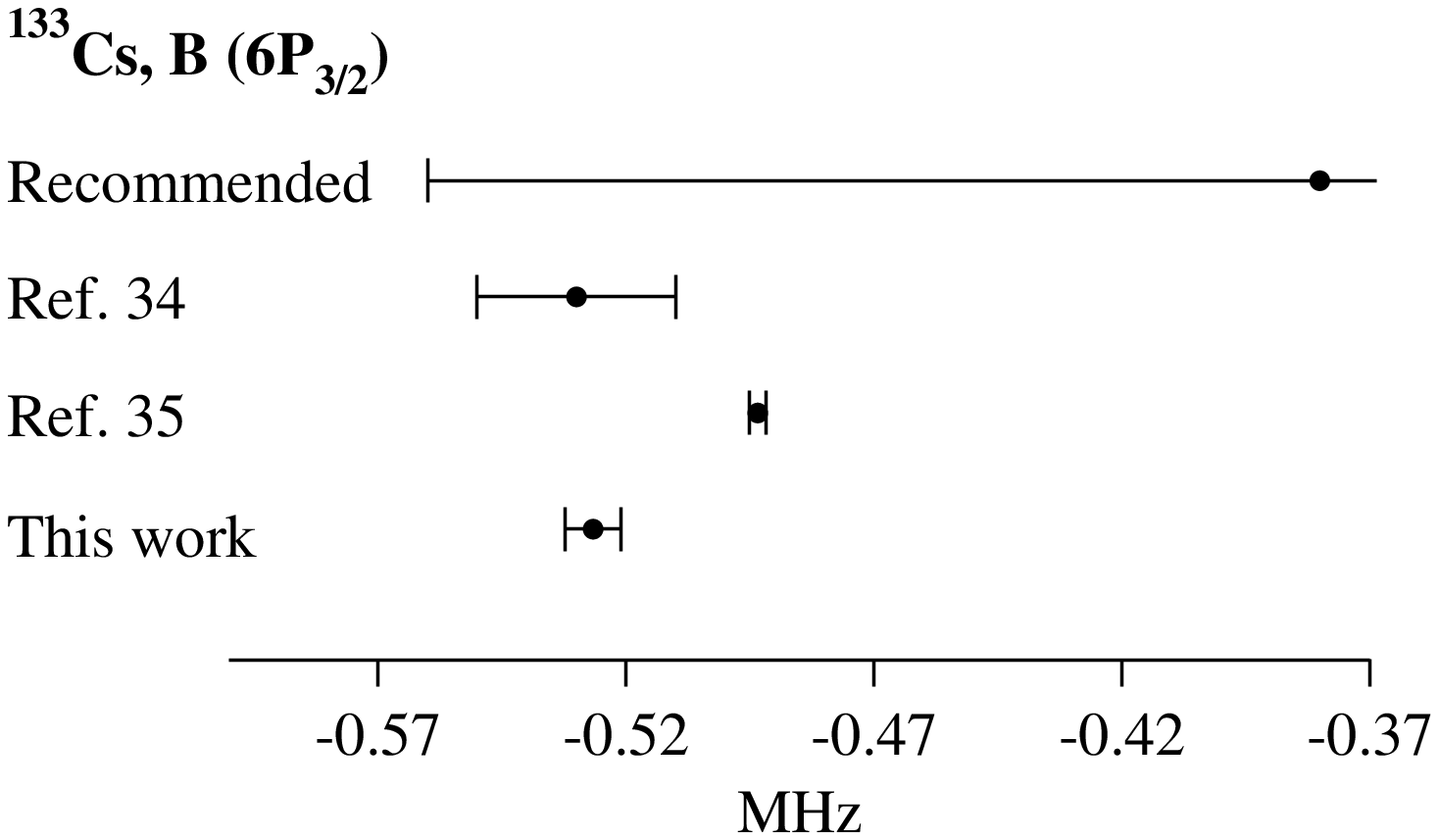}}
\caption{(a) Comparison of the value of $A$ in the
$6P_{1/2}$ state of $^{133}$Cs obtained in this work to
earlier values. (b) Comparison of the values of $A$ and $B$
in the $6P_{3/2}$ state of $^{133}$Cs obtained in this work
to earlier values.} \label{f6ab}
\end{figure}

Since the publication of our result, there has been a
report of another high-precision measurement of the
interval in this state by Gerginov {\it et al}.\
\cite{GCT06}. In that work, the absolute frequencies of the
transitions were measured using a femtosecond frequency
comb with a stated precision of 4 kHz. Our value is
compared to this value and other published values in Fig.\
\ref{f6ab}. Two of the previous values are also from
measurements of the absolute frequencies, one using a
frequency comb by H\"ansch and coworkers \cite{URH99}, and
the other from our laboratory using a Rb-stabilized
ring-cavity resonator \cite{DBB06}. Our value is consistent
with both these measurements and an earlier less precise
measurement in Ref.\ \cite{RAT97}, and appears inconsistent
only with the most recent measurement by Gerginov {\it et
al}.\ \cite{GCT06}.

\subsection{$D_2$ line}
The hyperfine levels in the $6P_{3/2}$ state are far apart
and the individual transitions in the $D_2$ line are well
resolved. However, the measurements on the $D_2$ line were
done using our technique of CCS and not SAS. The advantages
of the CCS technique for hyperfine measurements have been
highlighted earlier. In Cs, there is an additional
advantage because, in normal SAS, the hyperfine peaks can
get distorted and even change sign due to the effects of
optical pumping and radiation pressure \cite{SKW94}. On the
other hand, the lineshape in CCS remains symmetric even at
high powers. The measurements in Cs have been published
previously \cite{DAN05}. The large nuclear spin in
$^{133}$Cs implies that, at sufficiently high precision,
the intervals will show the effect of a small
magnetic-octupole interaction (characterized by the
hyperfine constant $C$). The average values of the
hyperfine intervals in terms of these constants are:
\begin{tabbing}
\hspace{0.75cm} $^{133}$Cs, $6P_{3/2}$: \=
$\Delta \nu_{5 - 4} = 5A + \frac{5}{7}B + \frac{40}{7}C$ \= $= 251.037(6)$ MHz, \\
\> $\Delta \nu_{4 - 3} = 4A - \frac{2}{7}B - \frac{88}{7}C$ \= $= 201.266(6)$ MHz, \\
\> $\Delta \nu_{3 - 2} = 3A - \frac{5}{7}B + \frac{88}{7}C$ \= $= 151.232(6)$ MHz.
\end{tabbing}

One of the motivations for our work on the $6P_{3/2}$ state
of Cs was that the published values of the hyperfine
interval $\Delta \nu_{5 - 4}$ had changed from 251.000(20)
MHz in the work of Tanner and Wieman \cite{TAW88} to
251.092(2) MHz in the recent work of Gerginov, Derevianko,
and Tanner \cite{GDT03}, a change of 92 kHz ($4.6 \sigma$).
This discrepancy is seen clearly in the deviation plot of
the hyperfine constants in Fig.\ \ref{f6ab}, comparing our
values to these previous results. Our results are close to
the earlier values \cite{TAW88}, but disagree with the
recent measurement by Gerginov {\it et al}.\ \cite{GDT03}.
It is also from a measurement by Gerginov {\it et al}.\
\cite{GCT06} that our interval in the $D_1$ line differs by
70 kHz.

\section{Summary}
In conclusion, we have recently developed a technique for
measuring hyperfine intervals using an AOM whose frequency
is directly locked to the frequency difference between two
transitions. We use techniques of high-resolution laser
spectroscopy to resolve hyperfine transitions in the $D$
lines of all the alkali atoms. In lithium, we study the
fluorescence signal from a collimated atomic beam excited
by a laser beam at right angles. In all other atoms, we
study the absorption of a probe beam through a vapor cell,
in the presence of either a counter-propagating pump beam
(saturated-absorption spectroscopy) or a co-propagating
control beam (coherent-control spectroscopy). We are thus
able to measure hyperfine intervals in the first-excited
$P_{1/2}$ and $P_{3/2}$ states of these atoms. This yields
the magnetic-dipole coupling constant $A$ and the
electric-quadrupole coupling constant $B$ in these states.
In the $6P_{3/2}$ state of $^{133}$Cs, there is evidence of
a small magnetic-octupole coupling constant $C$.

The measured hyperfine constants in the various atoms are
summarized in Table \ref{t5}. As discussed in earlier
sections, in most cases our values are consistent with
other measurements but have considerably improved
precision. Improved knowledge of the hyperfine constants
should prove useful to both theorists and experimentalists
working with alkali atoms. The only state in which we have
not measured hyperfine structure with higher precision is
the $2P_{3/2}$ state of Li, where the level spacing is
smaller than the natural linewidth.

In addition, we have previously used our Rb-stabilized
ring-cavity resonator to measure the absolute frequencies
of the $D$ lines in all atoms except $^{23}$Na. The
hyperfine intervals determined from the differences in the
transition frequencies match the intervals measured with
the direct AOM technique, though the precision is typically
five times less for the absolute-frequency method. The two
techniques are very different in the way the frequencies
are determined. Hence the consistency indicates that both
techniques are reliable within their error bars.

Of the 17 constants listed in the Table, there are only 2
values that are discrepant from recent measurements, namely
in the $P_{1/2}$ states of $^{39}$K and $^{133}$Cs. In both
cases, the recent measurements use the frequency-comb
technique to measure the absolute frequencies of the
transitions. However, in both cases there are other
independent measurements (albeit with slightly lower
precision) that agree with our results. In the case of
$^{39}$K, there is an earlier atomic beam magnetic
resonance measurement by Buck and Rabi \cite{BUR59}, and in
the case of $^{133}$Cs, there is a previous measurement by
H\"{a}nsch and coworkers also using the frequency-comb
technique \cite{URH99}.

Another common feature of the two frequency-comb
measurements is that they use fluorescence spectroscopy
from an atomic beam, while we use absorption spectroscopy
in a vapor cell. Cell experiments have two advantages over
the use of an atomic beam.
\begin{enumerate}
\item The line center is completely insensitive to the
first-order Doppler effect. Even if there were a small
misalignment angle between the pump/control and probe
beams, this would result in a broadening of the line and
not a shift. In a beam experiment, any deviation from
perpendicularity of the atomic beam from the laser beam
would result in a systematic Doppler shift.

\item The cell can be magnetically shielded much better
than a beam. Our residual field with the two-layer shield
is 1 mG, while the residual field in beam experiments is
typically 10 times larger. This could lead to
correspondingly larger Zeeman shifts.
\end{enumerate}
This is why we have done all our experiments with vapor
cells except in the case of lithium, for which it is
difficult to make vapor cells because hot lithium vapor is
highly reactive and attacks most glasses.

The frequency-comb technique is known to be highly
reliable. The AOM technique is also inherently free of
systematic errors at this level of precision since the rf
frequency can be measured with $< 0.1$ kHz accuracy. This
suggests that any discrepancy with frequency-comb
measurements arises due to errors in the atomic
spectrometers used, especially when the precision is being
pushed to $10^{-3}$ of the spectral linewidth. In other
words, the spectral linewidth and the ability to split the
line are going to limit the precision and not the
measurement technique {\it per se}. The discrepancy in the
interval for the $6P_{1/2}$ state of $^{133}$Cs between our
work and that of Gerginov {\it et al}.\ \cite{GCT06} is
about 70 kHz, or 1\% of the linewidth. It is conceivable
that one of the two spectrometers is systematically off by
this fraction of the linewidth. However, the discrepancy in
the interval for the $4P_{1/2}$ state of $^{39}$K from the
work of Falke {\it et al}.\ \cite{FTL06} is 2 MHz, or 33\%
of the linewidth. This is too large to be accounted for by
any known problems with the spectrometer. We have also done
the spectroscopy with both the SAS and the CCS techniques,
and get consistent results. Further high-precision
measurements with alternate techniques might shed light on
these two cases.

\ack We thank E.\ Arimondo for useful discussions related
to collisional shifts in hyperfine measurements. We are
grateful to W.~A. van Wijngaarden for help with the design
of the lithium oven and to Hema Ramachandran for loan of a
potassium vapor cell. This work was supported by the
Department of Science and Technology and the Board of
Research in Nuclear Sciences (DAE), Government of India.
One of us (D.D.) acknowledges a graduate fellowship from
the Council of Scientific and Industrial Research, India.

\section*{References}


\newpage

\begin{table}
\caption{Error budget.
\label{t1}}
\begin{indented}
\item[]
\begin{tabular}{lc}
\br
\hspace*{3mm} Source of error & Size (kHz) \\
\mr
1. Statistical error & 2 \\
2. Optical pumping into Zeeman sublevels & 3--5 \\
3. Feedback loop phase shift & 2 \\
4. Collisional shifts & 5 \\
\br
\end{tabular}
\end{indented}
\end{table}

\begin{table}
\caption{Measurements of hyperfine intervals in the ground
$2S_{1/2}$ state of Li. The second column lists the
transition to which the laser was locked, and the third
column lists the transition to which the AOM was locked.
Intensity-dependent errors were checked by repeating the
measurement at three powers, as listed in columns 4 to 6.
\label{t2}}
\begin{indented}
\item[]
\begin{tabular}{cccccc}
\br
Isotope & Laser & AOM & \multicolumn{3}{c}{Interval (MHz)} \\
\cline{4-6} & & & 15 $\mu$W & 28 $\mu$W & 42 $\mu$W \\
\mr $^{6}$Li & $\frac{3}{2} \rightarrow \frac{1}{2}$ &
$\frac{1}{2} \rightarrow \frac{1}{2}$
& 228.204  & 228.208 & 228.210 \\
& $\frac{3}{2} \rightarrow \frac{3}{2}$ & $\frac{1}{2} \rightarrow \frac{3}{2}$
& 228.207  & 228.211 & 228.205 \\
\\
$^{7}$Li
& $2 \rightarrow 1 $ & $1 \rightarrow 1$
& 803.506 & 803.514 & 803.511 \\
& $2 \rightarrow 2 $ & $1 \rightarrow 2$
& 803.507 & 803.508 & 803.513 \\
\br
\end{tabular}
\end{indented}
\end{table}

\begin{table}
\caption{Measurements at different powers of hyperfine
intervals in the excited $2P_{1/2}$ state of Li.
\label{t3}}
\begin{indented}
\item[]
\begin{tabular}{cccccc}
\br
Isotope & Laser & AOM & \multicolumn{3}{c}{Interval (MHz)} \\
\cline{4-6} & & & 15 $\mu$W & 28 $\mu$W & 42 $\mu$W \\
\mr $^{6}$Li & $\frac{3}{2} \rightarrow \frac{3}{2}$ &
$\frac{3}{2} \rightarrow \frac{1}{2}$
& 26.091 & 26.094 & 26.099 \\
& $\frac{1}{2} \rightarrow \frac{3}{2}$ & $\frac{1}{2} \rightarrow \frac{1}{2}$
& 26.096 & 26.098 & 26.098 \\
\\
$^{7}$Li
& $2 \rightarrow 2 $ & $2 \rightarrow 1$
& 92.051 & 92.055 & 92.055 \\
& $1 \rightarrow 2 $ & $1 \rightarrow 1$
& 92.048 & 92.052 & 92.054 \\
\br
\end{tabular}
\end{indented}
\end{table}

\begin{table}
\caption{Comparison of results for $A$ in the $2P_{1/2}$
state of $^{6,7}$Li to previous values. The last column
gives a theoretical calculation in $^{7}$Li. \label{t4}}
\begin{indented}
\item[]
\begin{tabular}{llc}
\br
\multicolumn{2}{c}{$A$ (MHz)} & Reference \\
\cline{1-2}
\multicolumn{1}{c}{$^{6}$Li} & \multicolumn{1}{c}{$^{7}$Li} & \\
\mr
17.394(4)  & 46.024(3) & This work \\
17.386(31) & 46.010(25) & \cite{WAC03} \\
17.48(15)  & 46.17(35) & \cite{RIT65} \\
17.375(18) & 45.914(25) & Recommended \cite{AIV77} \\
\multicolumn{1}{c}{--} & 45.945 & Theory \cite{GFJ01} \\
\br
\end{tabular}
\end{indented}
\end{table}

\begin{table}
\caption{Summary of measured hyperfine constants in the
different alkali atoms. In $^{133}$Cs, there is evidence of
a small magnetic-octupole constant $C = 0.87(32)$ kHz. FS:
fluorescence spectroscopy, SAS: saturated-absorption
spectroscopy, and CCS: coherent-control spectroscopy.
\label{t5}}
\begin{indented}
\item[]
\begin{tabular}{lccccc}
\br
Atom & State & $A$ (MHz) & $B$ (MHz) & & Technique \\
\mr
$^{6}$Li & $2P_{1/2}$ & 17.3940(40) & & & Beam (FS) \\
\\
$^{7}$Li & $2P_{1/2}$ & 46.0235(30) & & & Beam (FS) \\
\\
$^{23}$Na & $3P_{1/2}$ & 94.3485(50) & & & Cell (SAS) \\
 & $3P_{3/2}$ & 18.530(3) & 2.721(8) & & Cell (CCS) \\
\\
$^{39}$K & $4P_{1/2}$ & 28.848(5) & & & Cell (CCS) \\
 & $4P_{3/2}$ & 6.077(23) & 2.875(55) & & Cell (CCS) \\
\\
$^{85}$Rb & $5P_{1/2}$ & 120.645(5) & & & Cell (SAS) \\
 & $5P_{3/2}$ & 25.040\,3(11) & 26.008\,4(49) & & Cell (SAS) \\
\\
$^{87}$Rb & $5P_{1/2}$ & 406.119(7) & & & Cell (SAS) \\
 & $5P_{3/2}$ & 84.7200(16) & 12.4970(35) & & Cell (SAS) \\
\\
$^{133}$Cs & $6P_{1/2}$ & 291.913\,5(15) & & & Cell (SAS) \\
 & $6P_{3/2}$ & 50.281\,63(86) & $-0.526\,6(57)$ &  & Cell (CCS) \\
\br
\end{tabular}
\end{indented}
\end{table}

\begin{figure}
\hspace*{2.5cm}
\resizebox{0.55\columnwidth}{!}{\includegraphics{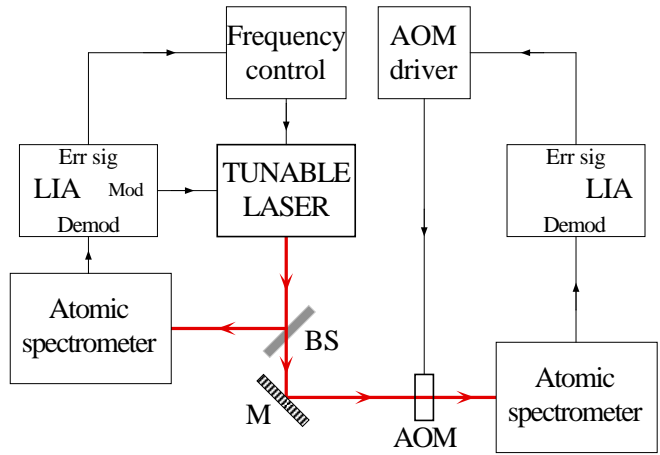}}
\caption{(Color online) Schematic of the experiment. Figure
key -- AOM: acousto-optic modulator, LIA: lock-in
amplifier, BS: beam splitter, M: mirror.} \label{f1}
\end{figure}

\begin{figure}
\hspace*{2cm}
\resizebox{0.43\columnwidth}{!}{\includegraphics{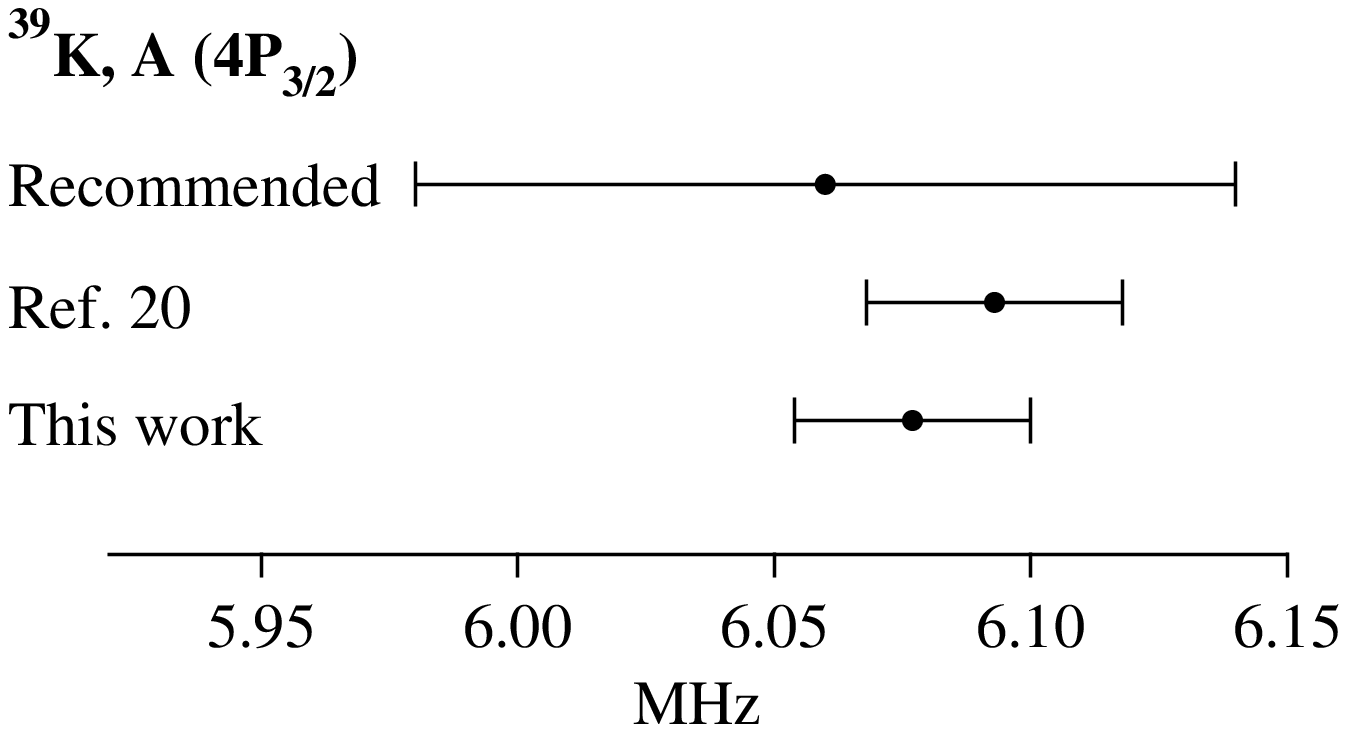}}
\resizebox{0.43\columnwidth}{!}{\includegraphics{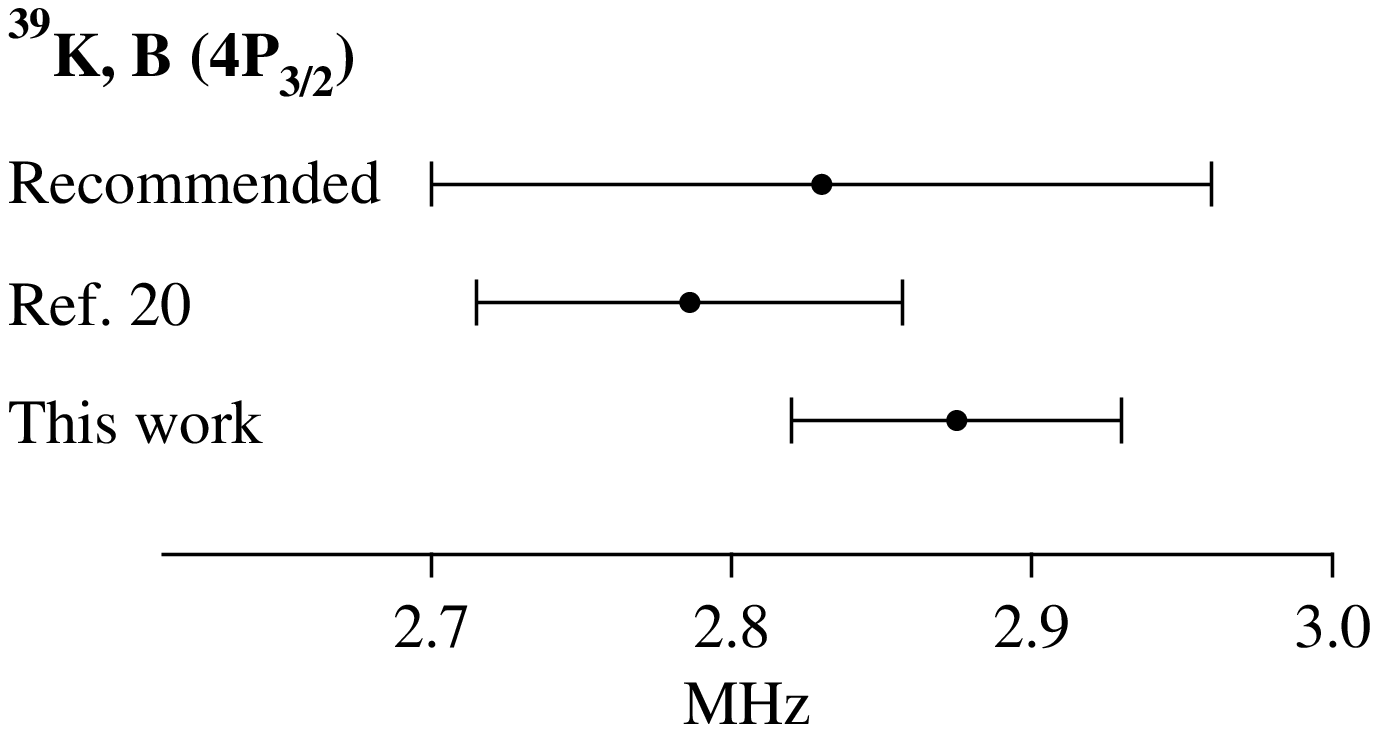}}
\caption{Comparison of the values of $A$ and $B$ in the
$4P_{3/2}$ state of $^{39}$K obtained in this work to
earlier values.} \label{f4d}
\end{figure}

\end{document}